\documentclass[12pt]{article}
\usepackage{epsf}
\usepackage[cp1251]{inputenc}
\usepackage[english]{babel}

\begin{document}

\title{
{\bf A simple model for high-energy nucleon-nucleon elastic diffraction and 
exclusive diffractive electroproduction of vector mesons on protons}}
\author{A.A. Godizov\thanks{E-mail: anton.godizov@gmail.com}\\
{\small {\it Institute for High Energy Physics, 142281 Protvino, Russia}}}
\date{}
\maketitle

\vskip-1.0cm

\begin{abstract}
The processes of exclusive diffractive scattering $p+p\to p+p$, $\bar p+p\to\bar p+p$, and 
$\gamma^*+p\to V+p$ at high energies are considered in the framework of a unified 
Regge-eikonal model with a very simple reggeon structure of the eikonal. It is demonstrated 
that the pomeron trajectory is universal in all reactions and having intercept about 1.31 
which could be extracted explicitly from the data on the proton structure function $F_2(x,Q^2)$. 
The predictions for the proton-proton cross-sections at LHC energies are given.
\end{abstract}

\vspace*{1cm}

\section*{Introduction}

The aim of this job is the construction of quite a simple phenomenological 
model which could be applicable to various processes of high energy diffractive scattering. 
For this purpose we will consider reactions of nucleon-nucleon elastic diffraction and 
exclusive electroproduction of vector mesons on protons and try to give a simultaneous 
description of these processes. This will be done in the framework 
of the Regge-eikonal approach \cite{collins} which allows to satisfy the Froissart-Martin bound 
\cite{froissart} explicitly. The basis of this approach is the eikonal (impact parameter) 
representation of the elastic diffractive non-flip scattering amplitude 
\begin{equation}
\label{eikrepr}
T_{12\to 12}(s,t) = 4\pi\lambda^{1/2}(s,m_1^2,m_2^2)\int_0^{\infty}db^2J_0(b\sqrt{-t})
\frac{e^{2i\delta_{12\to 12}(s,b)}-1}{2i}\,,
\end{equation}
where $s$ is the collision energy squared, $t$ is the transferred 4-momentum squared, $b$ is 
the impact parameter, $m_1$ and $m_2$ are the masses of the scattering particles, 
$\lambda(s,m_1^2,m_2^2)\equiv s^2+m_1^4+m_2^4-2m_1^2s-2m_2^2s-2m_1^2m_2^2$, 
and the eikonal is the sum of single-reggeon exchange terms \cite{collins} (for derivation 
of the eikonal Regge approximation see, also, Appendix): 
$$
\delta_{12\to 12}(s,b) = \frac{1}{16\pi\lambda^{1/2}(s,m_1^2,m_2^2)}\int_0^{\infty}d(-t)
J_0(b\sqrt{-t})\delta_{12\to 12}(s,t) = \frac{1}{16\pi\lambda^{1/2}(s,m_1^2,m_2^2)}\times
$$
\begin{equation}
\label{eikonal0}
\times\int_0^{\infty}d(-t)J_0(b\sqrt{-t})
\left\{\sum_n\left(i+{\rm tg}\frac{\pi(\alpha_n^+(t)-1)}{2}\right)
\Gamma_n^{(1)+}(t)\Gamma_n^{(2)+}(t)\left(\frac{s}{s_0}\right)^{\alpha_n^+(t)}\mp\right.
\end{equation}
$$
\left.\mp\sum_n\left(i-{\rm ctg}\frac{\pi(\alpha_n^-(t)-1)}{2}\right)
\Gamma_n^{(1)-}(t)\Gamma_n^{(2)-}(t)\left(\frac{s}{s_0}\right)^{\alpha_n^-(t)}\right\}\,.
$$
Here $\alpha_n^+(t)$, $\Gamma_n^{(i)+}(t)$ and $\alpha_n^-(t)$, $\Gamma_n^{(i)-}(t)$ are $C$-even and 
$C$-odd Regge trajectories and reggeon form-factors of the scattered particles, 
$s_0\equiv$ 1 GeV$^2$, and the sign ``$-$'' (``$+$'') before $C$-odd contributions corresponds to the 
particle-particle (particle-antiparticle) scattering. 

The practical use of the Regge-eikonal approach is that in the case of high energy diffraction 
it allows to reduce the unknown function of two variables, $T_{12\to 12}(s,t)$, to a few 
functions of one dynamical variable $t$ (Regge trajectories and reggeon form-factors) and 
to make explicit estimations for the high energy evolution of the diffractive pattern. 
In the following sections we will demonstrate that for various diffractive reactions 
there exist wide kinematical ranges where for description of cross-sections it is 
enough to keep only two $C$-even reggeons in the eikonal, pomeron and $f_2$-reggeon. 

Since the behavior of Regge trajectories and reggeon form-factors at low values of $t$ is still 
not calculable in the framework of QCD we have to use for them purely phenomenological 
(test) expressions. 
But though the main criterion for our choice of such parametrizations is simplicity we should 
take into account the QCD asymptotic behavior of Regge trajectories at $t\to-\infty$. For example, 
if we assume that in the limit of large transfers the pomeron exchanges turn into multi-gluon 
exchanges then (like in QED \cite{wu}) we come to \cite{low,kearney,kirschner} 
\begin{equation}
\label{gluon}
\lim_{t\to -\infty}\alpha_{\rm P}(t) = 1\,.
\end{equation}
In the case of the quark-antiquark pair ($f_2$-reggeon, {\it etc.}) one 
obtains \cite{kwiecinski} 
\begin{equation}
\label{meson}
\lim_{t\to -\infty}\alpha_{\bar qq}(t)\ln^{1/2}(-t)=\sqrt{\frac{32}{11N_c-2n_f}}=\sqrt{\frac{32}{21}}\;,
\end{equation}
where $N_c=3$ is the number of colors and $n_f=6$ is the full number of quark flavors.

The main feature of the proposed model is the use of universal (``universality'' means 
``uniqueness'', {\it i.e.} that corresponding reggeons are the same in applications of the model to various 
reactions) and 
essentially nonlinear Regge trajectories with asymptotical behavior satisfying (\ref{gluon}), (\ref{meson}).

\section*{The model for the nucleon-nucleon elastic diffraction}

Thus, for the nucleon-nucleon elastic diffraction we will exploit the following approximation to 
the eikonal in the momentum representation: 
\begin{equation}
\label{eikphen}
\delta(s,t) = \delta_{\rm P}(s,t)+\delta_f(s,t)=
\left(i+{\rm tg}\frac{\pi(\alpha_{\rm P}(t)-1)}{2}\right)
{\Gamma^{({\rm pp})}_{\rm P}}^2(t)\left(\frac{s}{s_0}\right)^{\alpha_{\rm P}(t)}+
\end{equation}
$$
+\left(i+{\rm tg}\frac{\pi(\alpha_f(t)-1)}{2}\right)
{\Gamma^{({\rm pp})}_f}^2(t)\left(\frac{s}{s_0}\right)^{\alpha_f(t)}.
$$

For the pomeron and the $f_2$-reggeon trajectories we choose the simplest 
parametrizations\footnote{One should not consider analytical properties of test parametrizations 
seriously. True Regge trajectories and reggeon form-factors 
have much more complicated analytical structure. Nevertheless, at 
negative values of the argument they could be approximated by simple monotonic test functions.} 
satisfying asymptotical conditions (\ref{gluon}), (\ref{meson}): 
\begin{equation}
\label{pomeron}
\alpha_{\rm P}(t) = 1+C_{\rm P}e^{B_{\rm P}t},\;\;\;
\alpha_f(t) = \sqrt{\frac{32}{21}}\ln^{-1/2}\frac{t_f-t}{\Lambda^2_f}\,.
\end{equation}
The reggeon test form-factors are chosen as 
\begin{equation}
\label{resid}
\Gamma^{({\rm pp})}_{\rm P}(t) = \Gamma^{({\rm pp})}_{\rm P}e^{b^{({\rm pp})}_{\rm P}t},\;\;\;\;
\Gamma^{({\rm pp})}_f(t) = \Gamma^{({\rm pp})}_f e^{b^{({\rm pp})}_f t}\,.
\end{equation}

To obtain angular distributions one should substitute (\ref{pomeron}), (\ref{resid}) into 
(\ref{eikphen}), then using (\ref{eikonal0}) and (\ref{eikrepr}) calculate the scattering 
amplitude (integrals are calculated numerically) and substitute it into the 
expression for the differential cross-section 
\begin{equation}
\label{diffsech}
\frac{d\sigma_{el}}{dt} = \frac{|T_{el}(s,t)|^2}{16\pi \lambda(s,m_p^2,m_p^2)}\,.
\end{equation}

\begin{figure}[ht]
\epsfxsize=8.2cm\epsfysize=8.2cm\epsffile{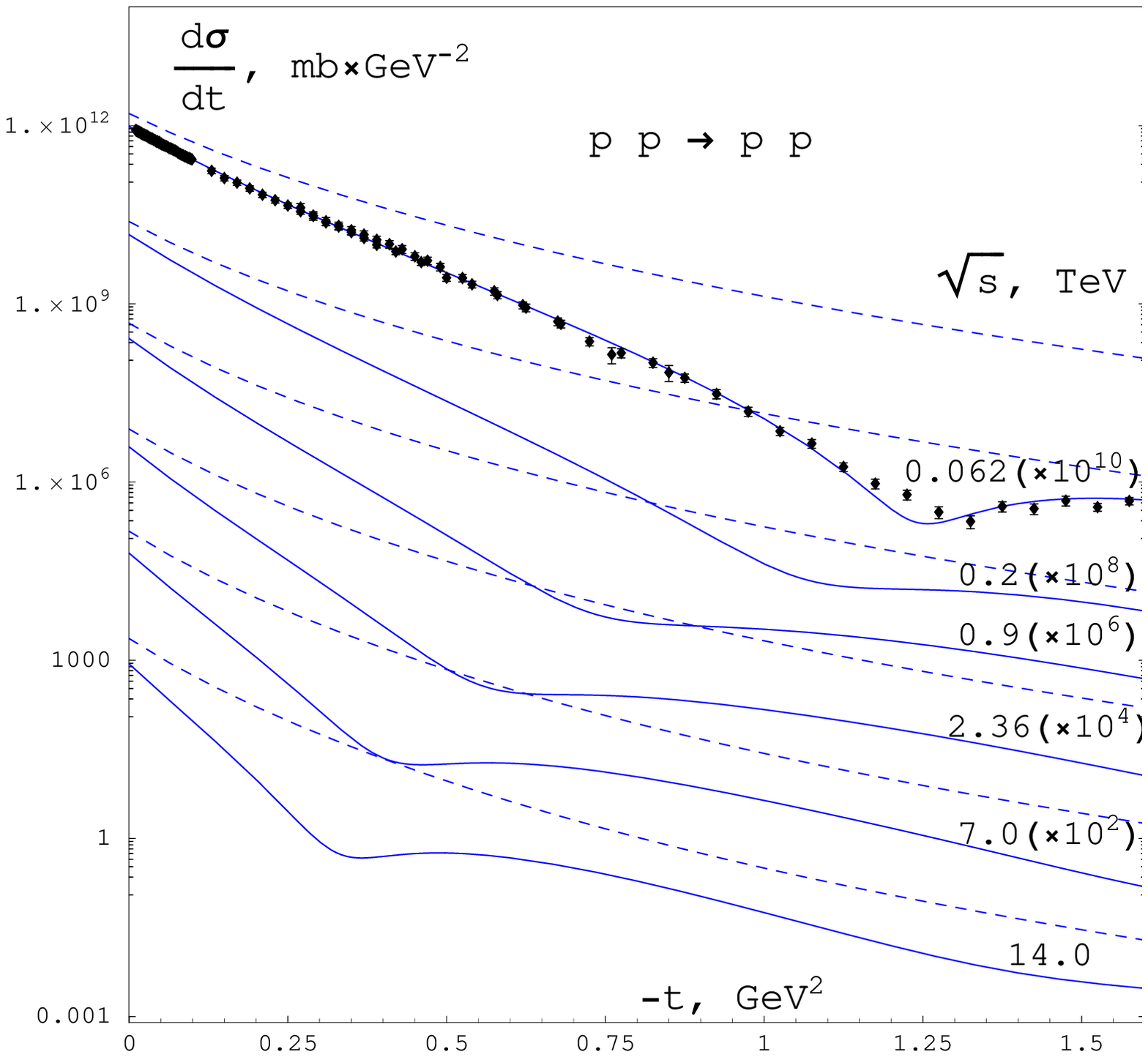}
\vskip -8.2cm
\hskip 8.5cm
\epsfxsize=8.2cm\epsfysize=8.2cm\epsffile{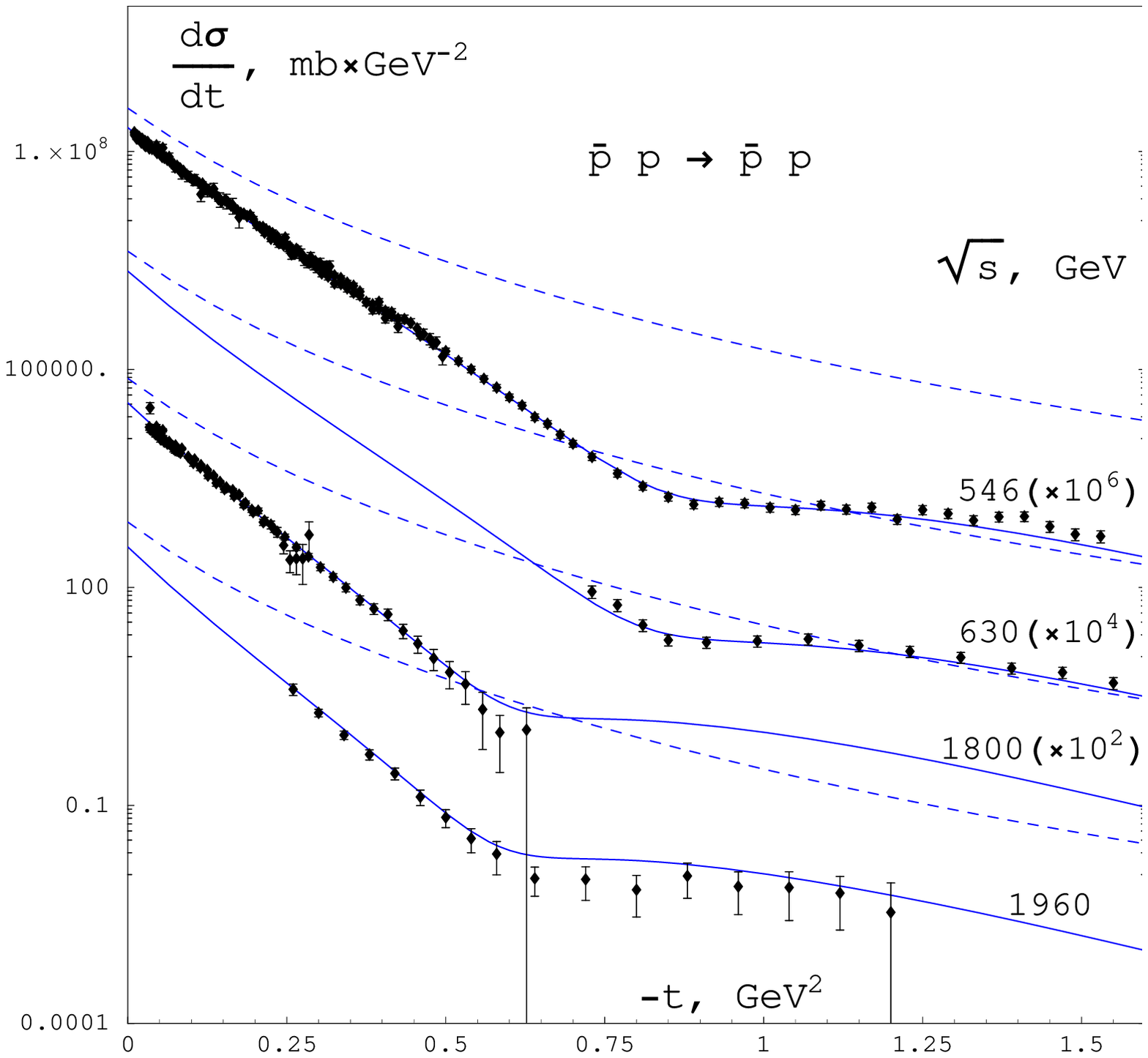}
\end{figure}
\begin{figure}[ht]
\vskip -0.85cm
\hskip 0.5cm
\epsfxsize=7.75cm\epsfysize=7.75cm\epsffile{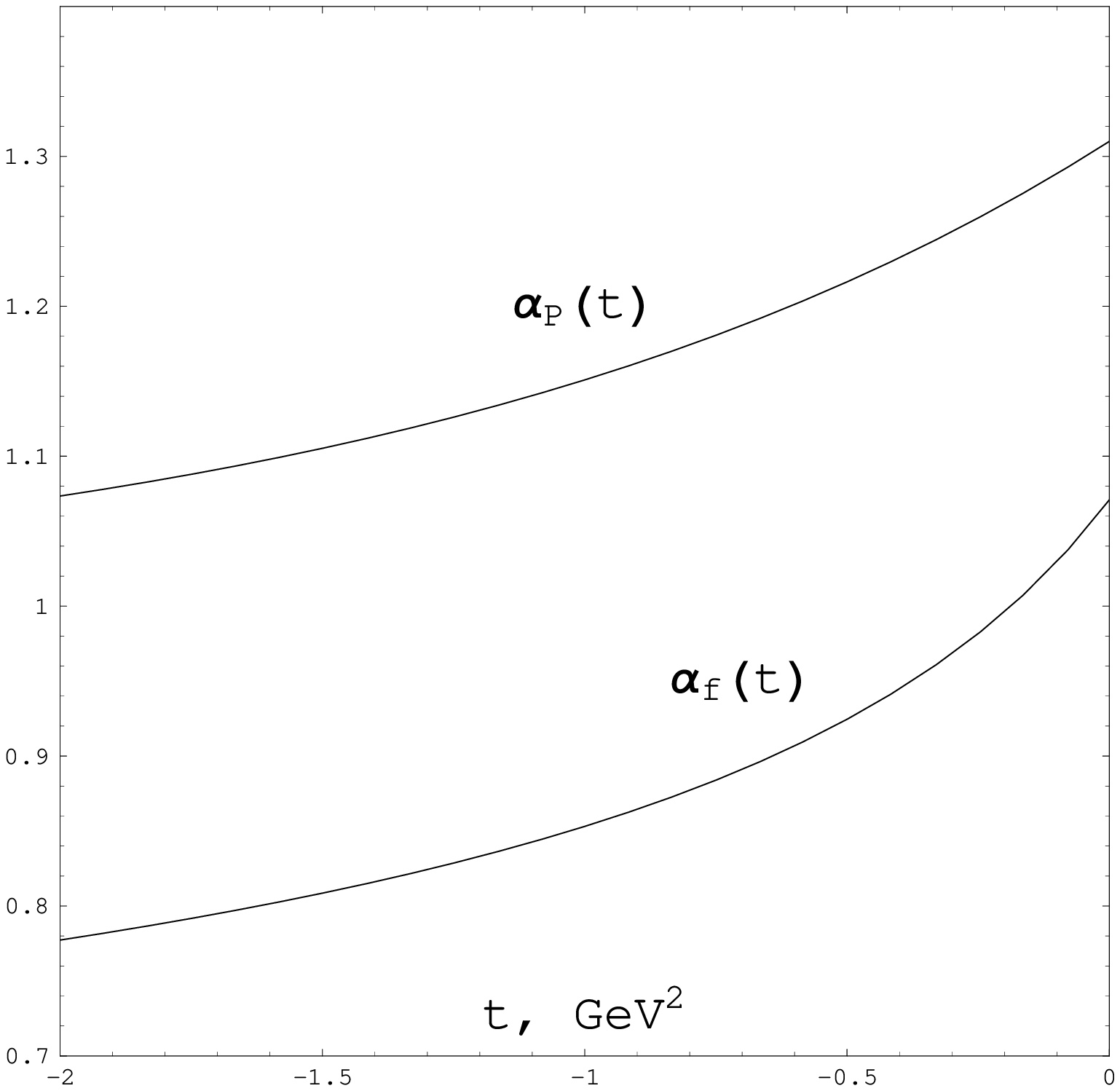}
\vskip -7.75cm
\hskip 8.95cm
\epsfxsize=7.75cm\epsfysize=7.75cm\epsffile{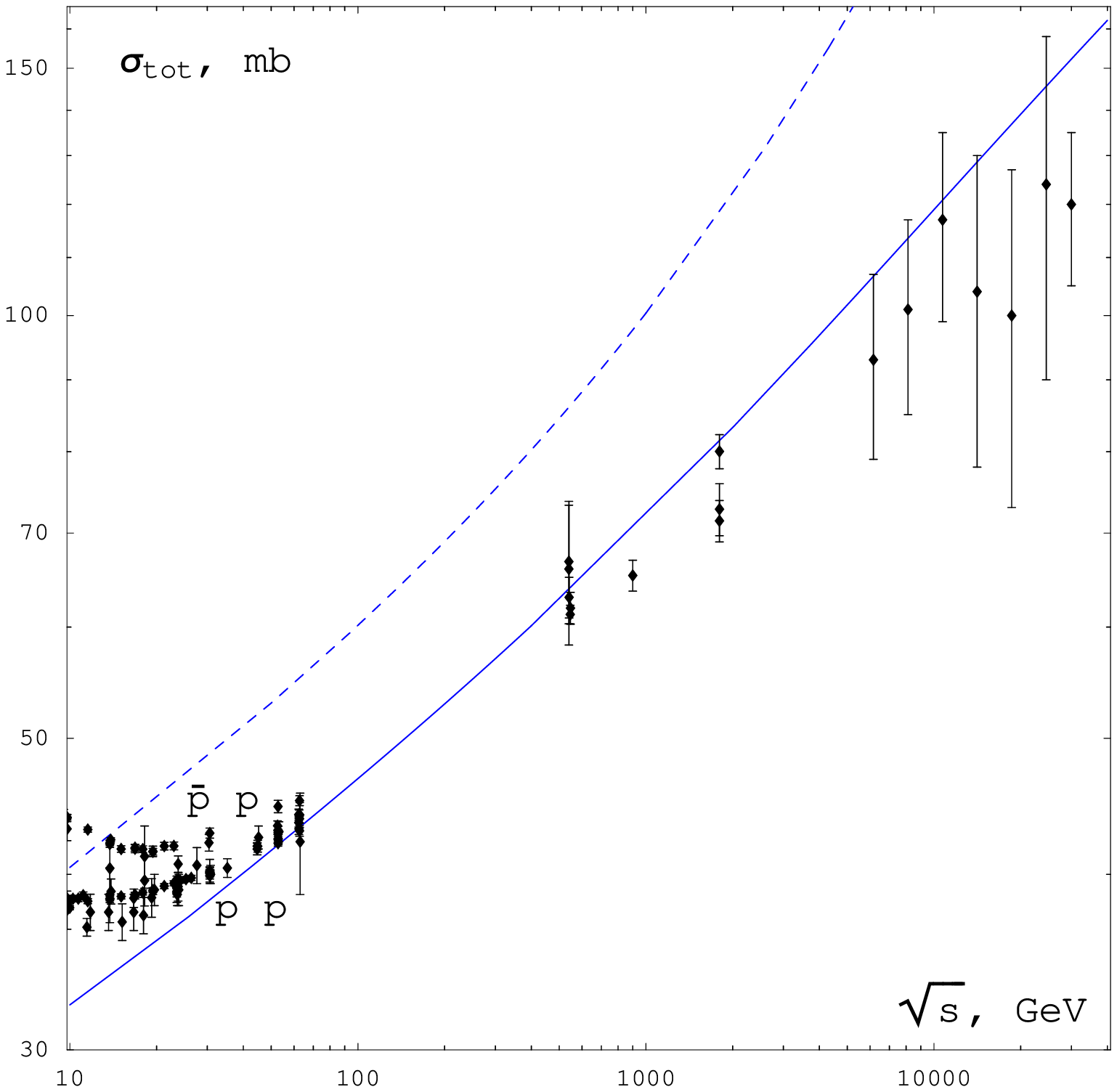}
\caption{Nucleon-nucleon cross-sections at different values of the collision energy and 
approximate Regge trajectories of the pomeron and the $f_2$-reggeon. The dashed lines correspond 
to the Born amplitudes.}
\label{pp}
\end{figure}

We postpone the discussion of the used parametrizations to the last section 
and now apply the proposed phenomenological scheme to the description of the elastic 
nucleon-nucleon diffractive scattering. Here we restrict ourselves by the kinematical range 
$\sqrt{s}>$ 60 GeV, 0.01 GeV$^2<-t<$ 1.6 GeV$^2$. 
At larger transfers the systematic deviations of the model curves from experimental angular 
distributions 
become too large due to the fact that exponential approximations to reggeon form-factors are 
invalid in the hard scattering region. 
At $\sqrt{s}<$ 60 GeV the contributions from secondary reggeons ($\omega$, $\rho$, $a$, {\it etc.}) 
have so significant influence on the values of dynamical 
quantities at low $t$ and in the region of the first dip that they also 
should not be ignored.\footnote{In fact, relative deviations about even one percent 
from the experimental data are noticeable in the near-forward scattering vicinity. So, we take into 
consideration only $pp$-scattering at $\sqrt{s}\approx$ 62 GeV since at such energy the  
influence of secondary reggeons is minimal in comparison with other ISR data sets, also
due to the fact that $C$-odd and $C$-even secondary contributions to the eikonal have opposite signs 
and compensate one another.} And, at last, 
in the region $\sqrt{-t}<$ 0.1 GeV the interference with Coulomb interaction takes place. 

The value of the pomeron intercept can be extracted explicitly from the data on $\gamma^*p$ total 
cross-sections at high photon virtualities and turns out $\approx 1.31$. We will 
discuss this below in one of the following sections. 

For verification of the model we used the set of data on angular distributions collected by 
J.R. Cudell, A. Lengyel, and E. Martynov at \cite{cudell}. The original data can also 
be found at \cite{diffexp}. The data set for $\sqrt{s}=1.96$ TeV is taken from \cite{d0}.
The results of applying the model to the data are represented in Fig. \ref{pp} and Tabs. 
\ref{tab1}, \ref{tab2}. 

\begin{table}[ht]
\begin{center}
\begin{tabular}{|l|l|l|l|}
\hline
\multicolumn{2}{|c|}{\bf pomeron}      &  \multicolumn{2}{|c|}{\bf $f$-reggeon}  \\
\hline
$C_{\rm P}$      & 0.31 (FIXED)              & $t_f$           & 0.87 GeV$^2$          \\
$B_{\rm P}$      & 0.72 GeV$^{-2}$      & $\Lambda_f$     & 0.48 GeV             \\
$\Gamma^{({\rm pp})}_{\rm P}$ & 0.95                 & $\Gamma^{({\rm pp})}_f$    & 8.5              \\
$b^{({\rm pp})}_{\rm P}$      & 0.24  GeV$^{-2}$    & $b^{({\rm pp})}_f$         & 1.15 GeV$^{-2}$   \\
\hline
\end{tabular}
\end{center}
\caption{Parameter values for test parametrizations of leading Regge 
trajectories and corresponding reggeon form-factors of nucleons.}
\label{tab1}
\end{table}

\begin{table}[ht]
\begin{center}
\begin{tabular}{|l|l|l|}
\hline
Set of data       & Number of points &  $\chi^2$        \\
\hline
$\sqrt{s}=$ 62   GeV ($p\,p$)        & 122    & 112     \\
$\sqrt{s}=$ 546  GeV ($\bar p\,p$)   & 221    & 291     \\
$\sqrt{s}=$ 630  GeV ($\bar p\,p$)   & 13     & 13      \\
$\sqrt{s}=$ 1800 GeV ($\bar p\,p$, E-710)   & 51     & 68     \\
$\sqrt{s}=$ 1800 GeV ($\bar p\,p$, CDF)   & 26     & 77     \\
$\sqrt{s}=$ 1960 GeV ($\bar p\,p$, D0)   & 17     & 25      \\
\hline
Total & 450 & 586     \\
\hline
\end{tabular}
\end{center}
\caption{The quality of description of the data on angular distributions of elastic 
nucleon-nucleon scattering.}
\label{tab2}
\end{table}

Also, in Fig. \ref{pp} the total cross-section dependence on the collision 
energy \cite{total} is given. In particular, $\sigma^{pp}_{tot}$(7 TeV) $\approx$ 110 mb and 
$\sigma^{pp}_{tot}$(14 TeV) $\approx$ 128 mb. 

\section*{The model for the exclusive diffractive electroproduction of vector mesons on protons}

We will treat exclusive diffractive electroproduction of vector mesons on protons from the 
standpoint of the vector-dominance model (VDM) \cite{sakurai} where the incoming photon 
fluctuates into a virtual vector meson which, in turn, scatters from the target proton. 
According to the so-called hypothesis of $s$-channel helicity conservation (the adequacy of 
this approximation is confirmed by numerous experimental data on helicity effects 
\cite{jpsiele,phiele}) the cross-section of reaction $\gamma^*+p\to V+p$ (here $V$ denotes 
some vector meson) is dominated by 
the non-flip helicity amplitude which can be represented in the form 
\begin{equation}
\label{genvdm}
T^\lambda_{\gamma^* p\to V p}(W^2,t,Q^2)=\sum_{V'}C_{V'}^\lambda(Q^2)
T^\lambda_{V'^* p\to V p}(W^2,t,Q^2)\,,
\end{equation}
where $W$ is the collision energy, $Q^2$ is the incoming photon virtuality, 
$T^\lambda_{V'^* p\to V p}(W^2,t,Q^2)$ is the diffractive hadronic amplitude with Regge-eikonal 
structure (the applicability of the Regge-eikonal approach to the hadronic reactions with off-shell 
particles is grounded in \cite{petrov}), and the sum 
is taken over all neutral vector mesons. At low $Q^2$ the function $C_{V'}^\lambda(Q^2)$ behaves like 
the coefficient of vector dominance (at high $Q^2$ the VDM is invalid \cite{jpsiele,phiele}) 
dependent on the type of vector meson $V'$, helicity $\lambda$, and virtuality $Q^2$. 

It is known that for small enough values of $t$ the diagonal coupling of vacuum reggeons to hadrons 
is much stronger than the off-diagonal coupling (for example, elastic proton-proton scattering has 
a considerably larger cross-section than diffractive excitation of N(1470) in the proton-proton 
collisions). Hence, in the high-energy range ($W>30$ GeV) where vacuum exchanges exceed the non-vacuum ones 
the so-called ``diagonal approximation'' of (\ref{genvdm}) may be used 
\begin{equation}
\label{diavdm}
T^\lambda_{\gamma^* p\to V p}(W^2,t,Q^2)=
C_{V}^\lambda(Q^2)T_{V^* p\to V p}(W^2,t,Q^2).
\end{equation}
Here we should note that measurements by ZEUS Collaboration \cite{jpsiele,phiele} did not reveal patent 
dependence on $t$ and $W$ of the ratio of differential cross-sections for the cases of longitudinally 
and transversely polarized incoming photon and this is the reason why we have omitted the superscript 
$\lambda$ for $T_{V^* p\to V p}(W^2,t,Q^2)$ in (\ref{diavdm}). In other words, we presume that spin 
phenomena related to the dependence of $T_{V^* p\to V p}(W^2,t,Q^2)$ on helicities of incoming and 
outgoing particles are much more fine effects than diffractive scattering itself and further deal only 
with amplitude $T_{\gamma^* p\to V p}(W^2,t,Q^2)$ averaged over spin states (it is 
obtained by the replacement $C_{V}^\lambda(Q^2)\to \sqrt{\frac{1}{3}\sum_\lambda {[C_{V}^\lambda(Q^2)]}^2}$ 
in (\ref{diavdm})).\footnote{The $Q^2$-behavior of $C_{V}^\lambda(Q^2)$ differs 
at different $\lambda$ but in this job we concentrate on the reggeon structure of the amplitude 
and, so, on the helicity independent $W$- and $t$-behavior of cross-sections.} For this quantity 
the extended (off-shell) eikonal representation is valid \cite{petrov}: 
\begin{equation}
\label{eikreprz}
T_{\gamma^* p\to V p}(W^2,b,Q^2) = \frac{\delta_{\gamma^* p\to V p}(W^2,b,Q^2)}
{\delta_{V p\to V p}(W^2,b)}T_{V p\to V p}(W^2,b) = 
\end{equation}
$$
=\delta_{\gamma^* p\to V p}(W^2,b,Q^2)+
i\delta_{\gamma^* p\to V p}(W^2,b,Q^2)\delta_{V p\to V p}(W^2,b)+...
$$
(where $b$ is the impact parameter, $\delta_{\gamma^* p\to V p}(W^2,b,Q^2)$ is the ``eikonal'' (the sum 
of pole terms) of the vector meson electroproduction on protons, $\delta_{V p\to V p}(W^2,b)$ 
is the eikonal of the vector meson elastic scattering on protons, and 
$T_{V p\to V p}(W^2,b)=\frac{e^{2i\delta_{V p\to V p}(W^2,b)}-1}{2i}$ is 
the ``eikonalized'' (unitarized) amplitude of elastic $Vp$-scattering). 

The secondary reggeon exchanges are suppressed 
($C$-odd reggeon exchanges -- due to the $C$-parity conservation, $a_2$-reggeon exchange -- due to the 
isospin conservation). Hence, the pole part of the amplitude takes a relatively simple form:
\begin{equation}
\label{poleconz}
\delta_{\gamma^* p\to V p}(W^2,t,Q^2) = \left[\left(i+{\rm tg}\frac{\pi(\alpha_{\rm P}(t)-1)}{2}\right)
\Gamma^{({\rm V^*V})}_{\rm P}(t,Q^2)\Gamma^{({\rm pp})}_{\rm P}(t)
\left(\frac{W^2}{W_0^2}\right)^{\alpha_{\rm P}(t)}
+\right.
\end{equation}
$$
\left.+\left(i+{\rm tg}\frac{\pi(\alpha_f(t)-1)}{2}\right)
\Gamma^{({\rm V^*V})}_f(t,Q^2)\Gamma^{({\rm pp})}_f(t)
\left(\frac{W^2}{W_0^2}\right)^{\alpha_f(t)}\right]
\sqrt{\frac{3\Gamma_{V\to e^+e^-}}{\alpha_e M_V}}
\frac{M^2_V}{M^2_V+Q^2}\;,
$$
where $M_V$ is the vector meson mass, $\alpha_e\approx\frac{1}{137}$ is the electromagnetic 
coupling, $\Gamma_{V\to e^+e^-}$ is the width of the vector meson decay to the electron-positron pair, 
$W_0\equiv 1$ GeV, $\alpha_{\rm P}(t)$ and $\alpha_f(t)$ are the pomeron and the $f_2$-reggeon trajectories, 
$\Gamma^{({\rm pp})}_{\rm P}(t)$ and $\Gamma^{({\rm pp})}_f(t)$ are the corresponding reggeon 
form-factors of the proton (the factor $\sqrt{\frac{3\Gamma_{V\to e^+e^-}}{\alpha_e M_V}}
\frac{M^2_V}{M^2_V+Q^2}$ is singled out explicitly from the photon-meson-reggeon vertex 
functions for convenience). 
\begin{figure}[ht]
\epsfxsize=8.2cm\epsfysize=8.2cm\epsffile{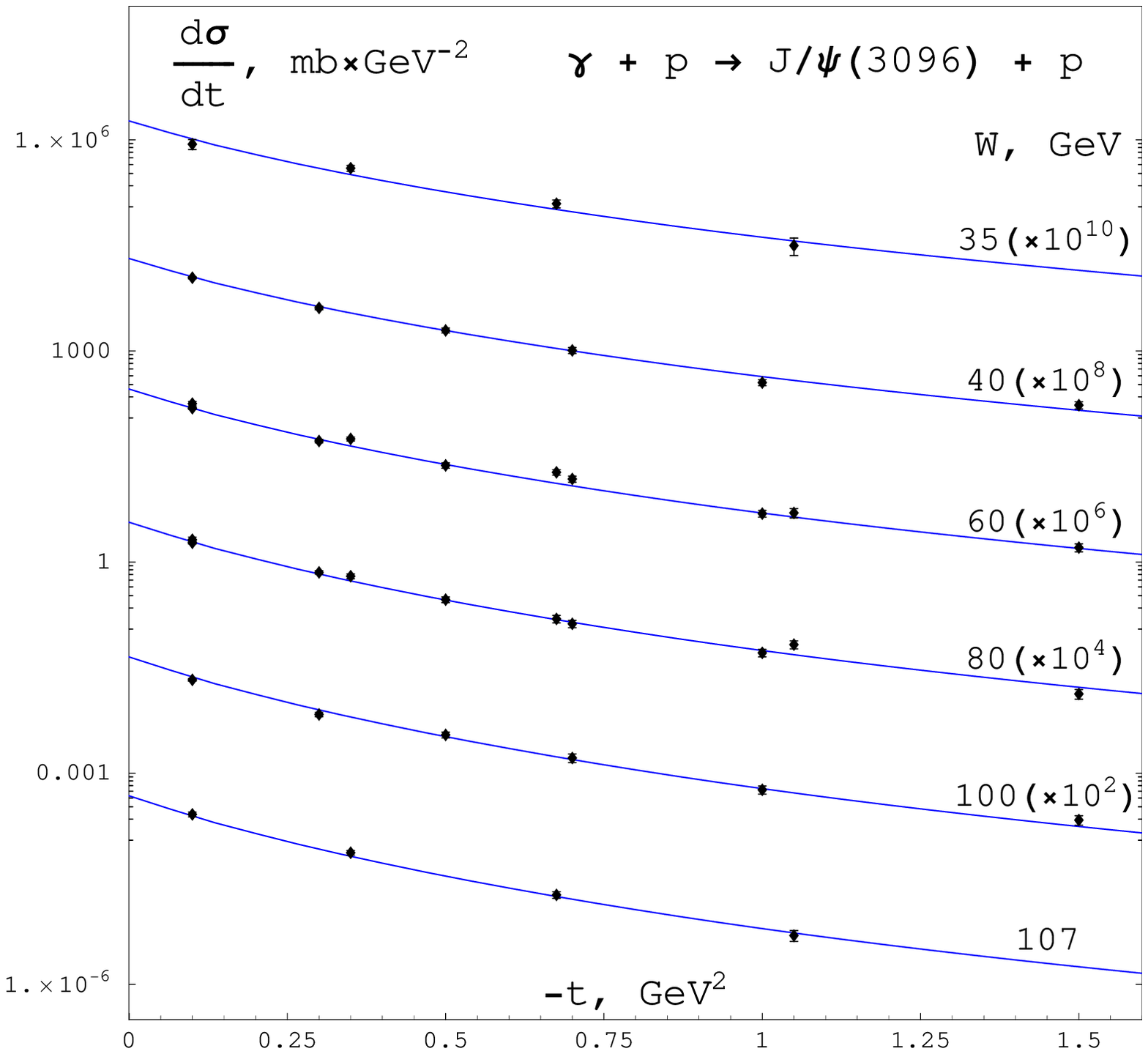}
\vskip -8.25cm
\hskip 8.5cm
\epsfxsize=8.2cm\epsfysize=8.2cm\epsffile{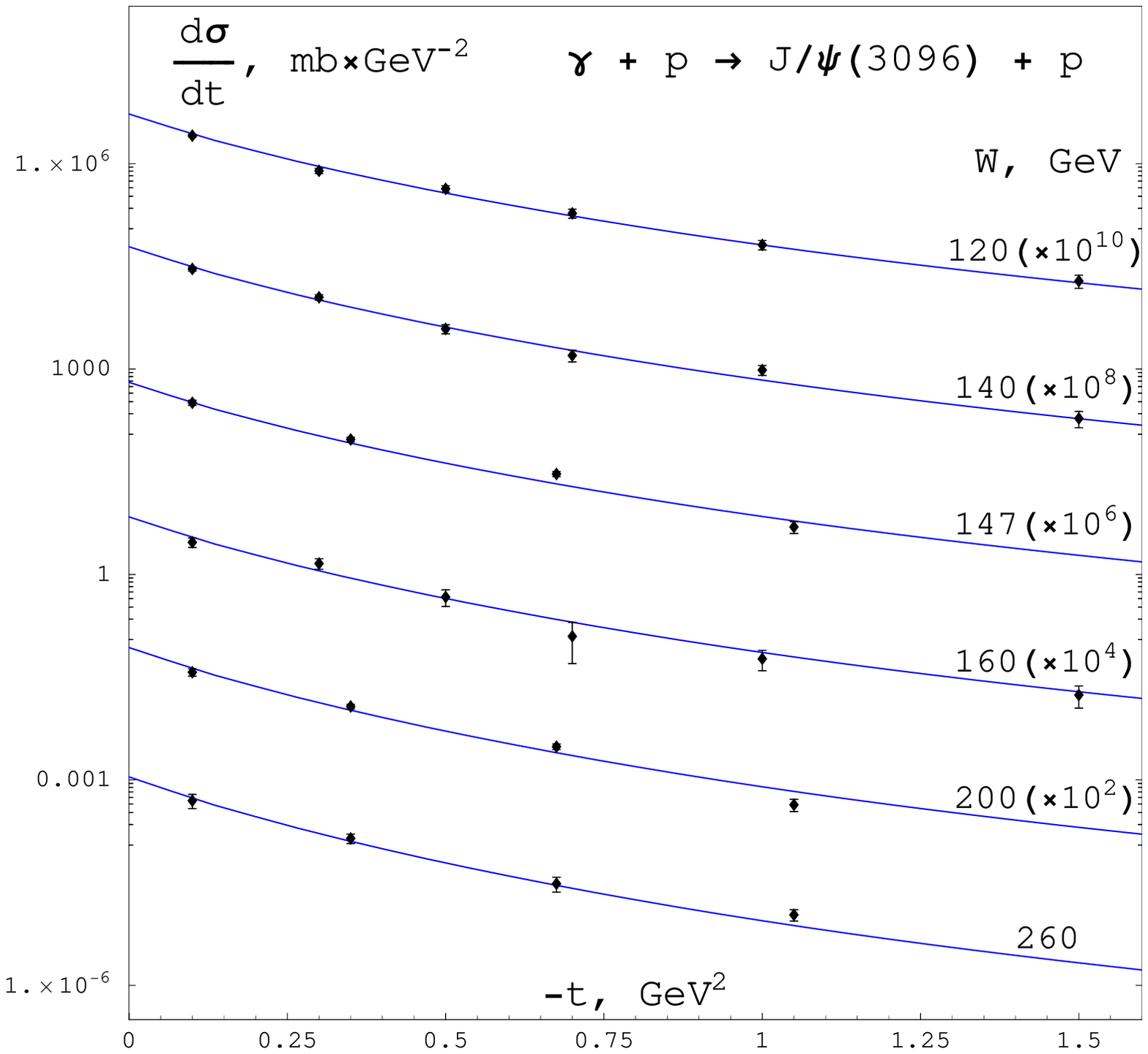}
\end{figure}
\begin{figure}[ht]
\vskip -0.85cm
\hskip 4.5cm
\epsfxsize=8.2cm\epsfysize=8.2cm\epsffile{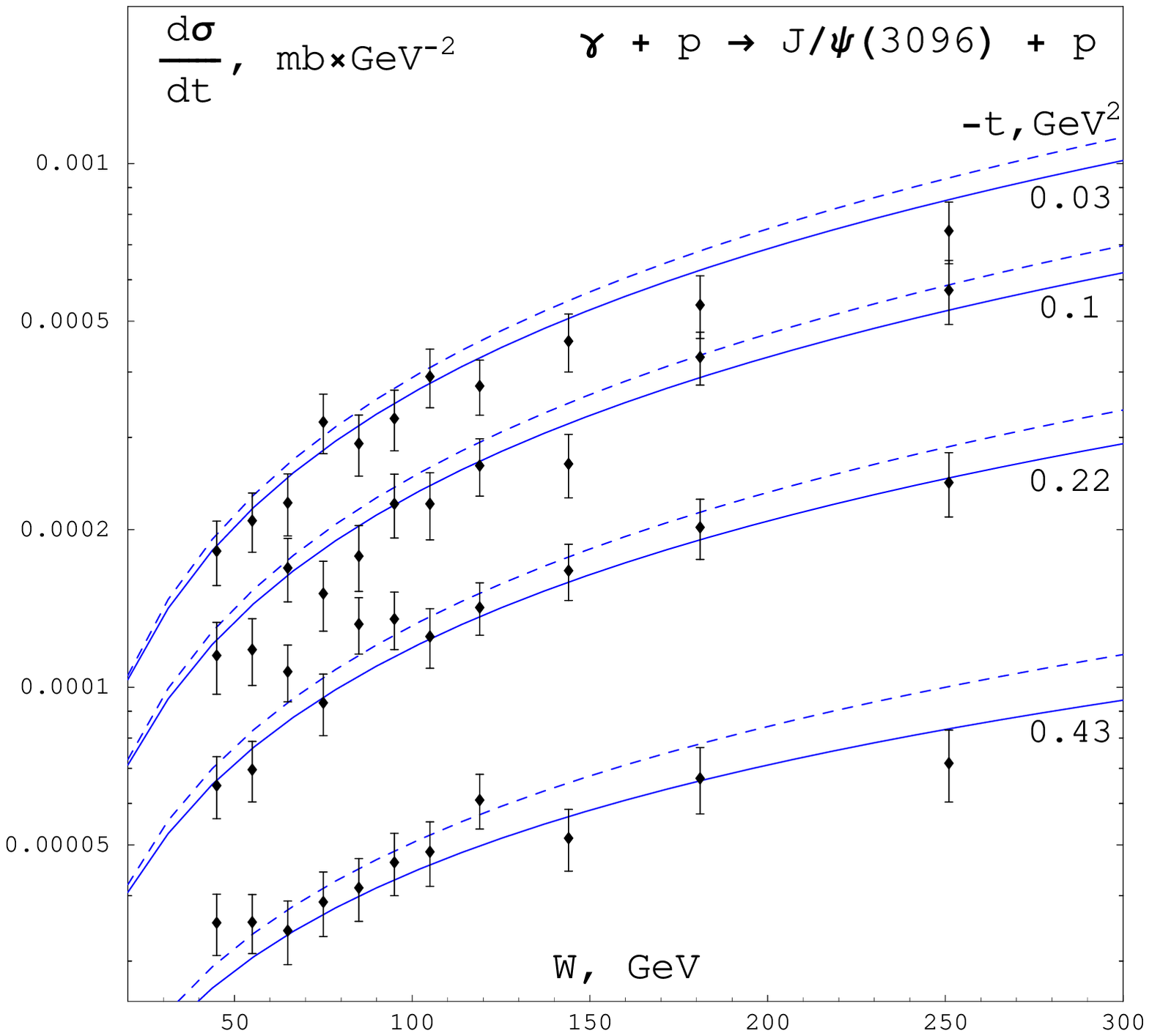}
\caption{Differential cross-sections of the $J/\psi(3096)$ photoproduction at different 
values of the collision energy. The dashed lines correspond to the Born amplitudes.}
\label{jps}
\end{figure}
The VDM implies that $\Gamma^{({\rm V^*V})}_{\rm R}(t,-M_V^2)=
\Gamma^{({\rm V \, V})}_{\rm R}(t)$, where $\Gamma^{({\rm V \, V})}_{\rm R}(t)$ is 
the reggeon (${\rm R}={\rm P},f$) form-factor of the vector meson:
\begin{equation}
\label{eikV}
\delta_{V p\to V p}(W^2,t) = \left(i+{\rm tg}\frac{\pi(\alpha_{\rm P}(t)-1)}{2}\right)
\Gamma^{({\rm V \, V})}_{\rm P}(t)\Gamma^{({\rm pp})}_{\rm P}(t)
\left(\frac{W^2}{W_0^2}\right)^{\alpha_{\rm P}(t)}
$$
$$
\left(i+{\rm tg}\frac{\pi(\alpha_f(t)-1)}{2}\right)
\Gamma^{({\rm V \, V})}_f(t)\Gamma^{({\rm pp})}_f(t)
\left(\frac{W^2}{W_0^2}\right)^{\alpha_f(t)}.
\end{equation}
To calculate $T_{\gamma^* p\to V p}(W^2,t,Q^2)$ we should determine functions 
$\Gamma^{({\rm V^*V})}_{\rm R}(t,Q^2)$ and $\Gamma^{({\rm V \, V})}_{\rm R}(t)$ 
(functions $\alpha_{\rm R}(t)$ and $\Gamma^{({\rm pp})}_{\rm R}(t)$ were fixed under consideration of 
elastic nucleon-nucleon scattering). 
\begin{figure}[ht]
\vskip -0.5cm
\epsfxsize=8.2cm\epsfysize=8.2cm\epsffile{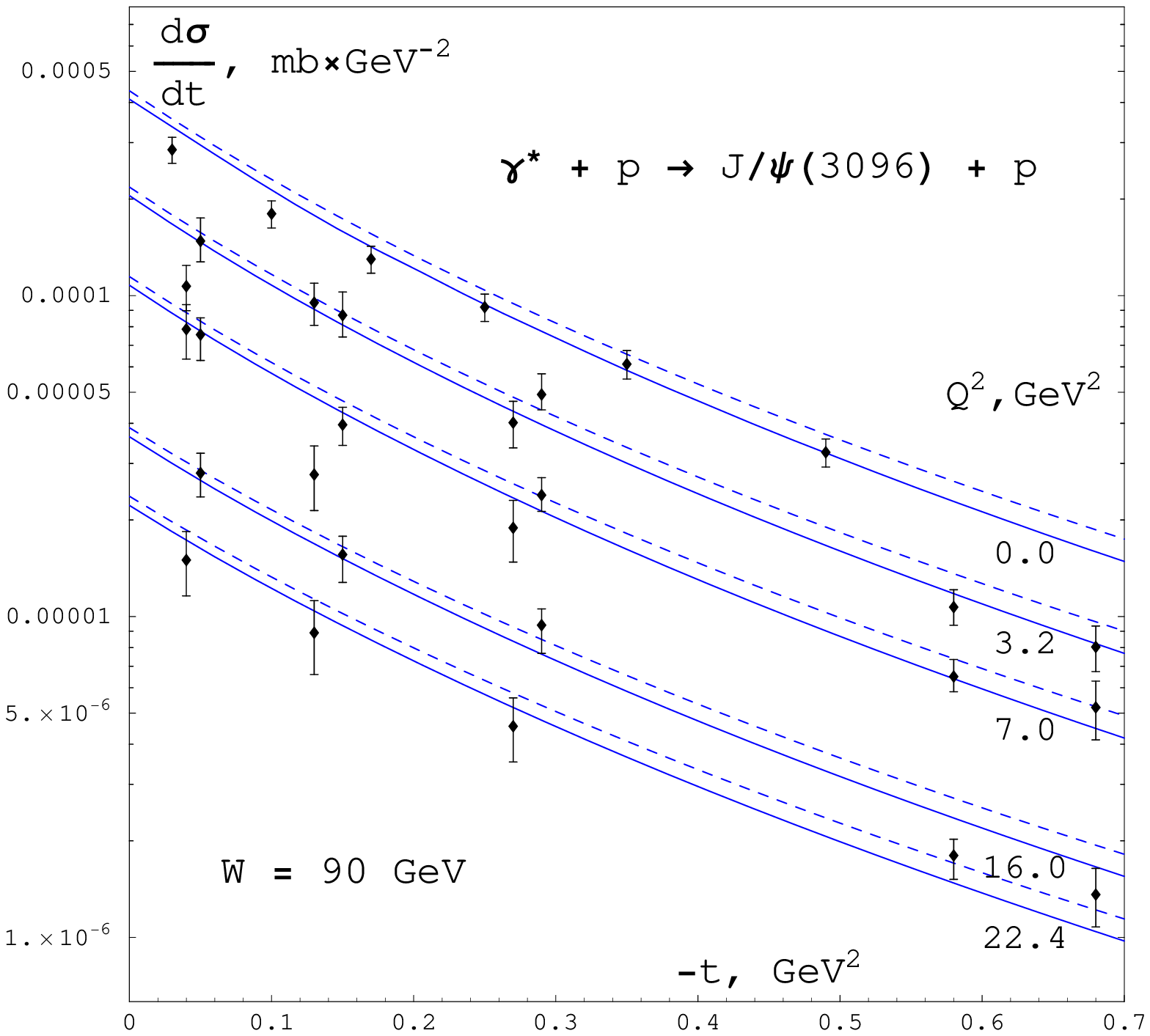}
\vskip -8.25cm
\hskip 8.5cm
\epsfxsize=8.2cm\epsfysize=8.2cm\epsffile{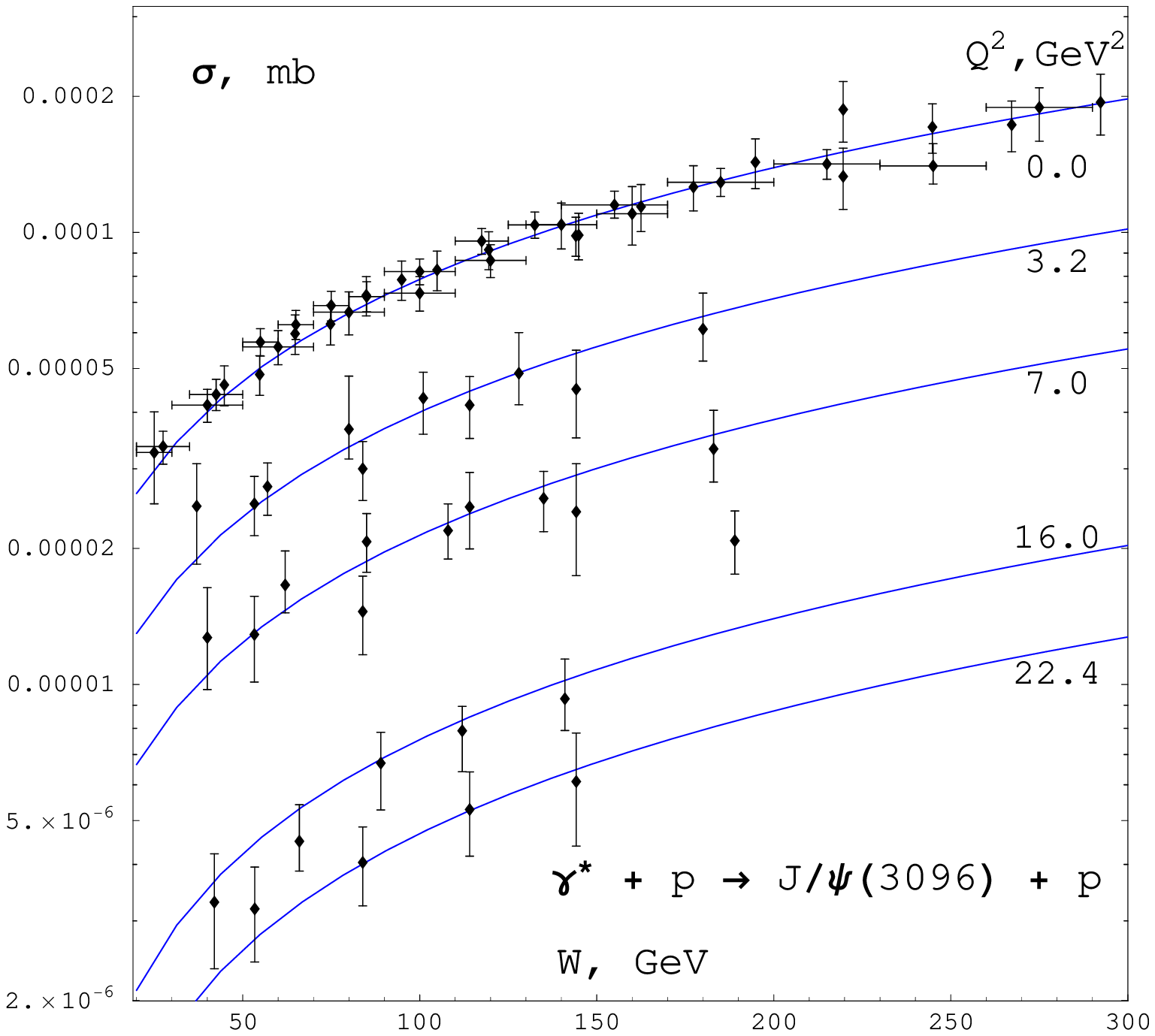}
\caption{Differential and integrated cross-sections of the $J/\psi(3096)$ electroproduction at different 
values of the incoming photon virtuality. The dashed lines correspond to the Born amplitudes.}
\label{jpsz}
\end{figure}

\begin{table}[ht]
\begin{center}

\begin{tabular}{|l|l|l|l|l|l|}
\hline
$Q^2$, GeV$^2$ & 0.0 & 3.2 & 7.0 & 16.0 & 22.4  \\
\hline
$\Gamma^{(J/\psi^*J/\psi)}_{\rm P}(Q^2)$ & 0.33 & 0.32 & 0.31 & 0.3 & 0.3   \\
\hline
$\Gamma^{(J/\psi^*J/\psi)}_f(Q^2)$ & 0.19 & 0.17 & 0.15 & 0.11 & 0.1   \\
\hline
\end{tabular}

\vskip 0.3cm

\begin{tabular}{|l|l|l|l|l|l|l|l|l|}
\hline
$Q^2$, $GeV^2$ & 0.0 & 2.4 & 3.5 & 5.0 & 6.6 & 9.2 & 13.0 & 19.7 \\
\hline
$\Gamma^{({\rm \phi^*\phi})}_{\rm P}(Q^2)$ & 0.6 & 1.5 & 1.6 & 1.6 & 1.6 & 1.6 & 1.6 & 1.4   \\
\hline
$\Gamma^{({\rm \phi^*\phi})}_f(Q^2)$ & 3.0 & 2.0 & 1.9 & 1.9 & 1.7 & 1.6 & 1.5 & 1.3   \\
\hline
\end{tabular}

\vskip 0.3cm

\begin{tabular}{|l|l|l|l|l|l|l|l|l|l|l|l|l|l|l|}
\hline
$Q^2$, $GeV^2$ & 0.0 & 2.5 & 3.5 & 5.0 & 6.0 & 6.6 & 8.0 & 11.7 & 13.5 & 17.4 & 19.7 & 33.0 & 35.6 & 41.0 \\
\hline
$\Gamma^{({\rm \rho^*\rho})}_{\rm P}(Q^2)$ & 0.41 & 1.5 & 1.6 & 1.7 & 1.8 & 1.8 & 2.0 & 2.0 & 1.95 & 1.9 & 1.9 & 1.8 & 1.8 & 1.7   \\
\hline
$\Gamma^{({\rm \rho^*\rho})}_f(Q^2)$ & 4.2 & 3.7 & 3.6 & 3.3 & 3.1 & 3.0 & 2.75 & 2.3 & 2.2 & 1.8 & 1.7 & 1.1 & 1.0 & 0.9   \\
\hline
\end{tabular}

\end{center}
\caption{Reggeon form-factors of vector mesons at different values of the incoming photon virtuality.}
\label{tab3}
\end{table}

Here we make an assumption that in the diffraction region one can neglect the $t$-dependence of reggeon 
form-factors of vector mesons ({\it i.e.} at low enough values of $t$ the approximations 
$\Gamma^{({\rm V^*V})}_{\rm R}(t,Q^2)\approx\Gamma^{({\rm V^*V})}_{\rm R}(Q^2)$ 
and $\Gamma^{({\rm V \, V})}_{\rm R}(t)\approx\Gamma^{({\rm V \, V})}_{\rm R}$ are valid\footnote{The 
hypothesis about the weak $t$-dependence of vector meson reggeon form-factors 
(in comparison with corresponding form-factors of the proton) is confirmed {\it a posteriori} under describing the 
experimental data.}). 
Such a behavior corresponds to the small effective radii of vector meson interaction with 
reggeons.
According to the dimensional counting rules \cite{matveev}, the quantities  
$\Gamma^{({\rm V^*V})}_{\rm R}(Q^2)$ should exhibit slow (nonpower-like) $Q^2$-behavior 
(power-like $Q^2$-dependence was already singled out above). 
Taking into account this slow $Q^2$-evolution we, also, assume that 
$\Gamma^{({\rm V \, V})}_{\rm R}\approx\Gamma^{({\rm V^*V})}_{\rm R}(0)$ 
(since we have no possibility to fix the on-shell form-factors separately from the 
elastic scattering of vector mesons on protons). Thus, for estimation of the photoproduction 
($Q^2\approx 0$) amplitude we need to determine only two extra free 
parameters, $\Gamma^{({\rm V^*V})}_{\rm P}(0)$ and $\Gamma^{({\rm V^*V})}_f(0)$. 

Applying the model to the data on the photoproduction of $J/\psi(3096)$ on protons \cite{jpsipho} yields 
$\chi^2\approx 158$ over 114 points (see Fig. \ref{jps}). 

\begin{figure}[ht]
\epsfxsize=8.2cm\epsfysize=8.2cm\epsffile{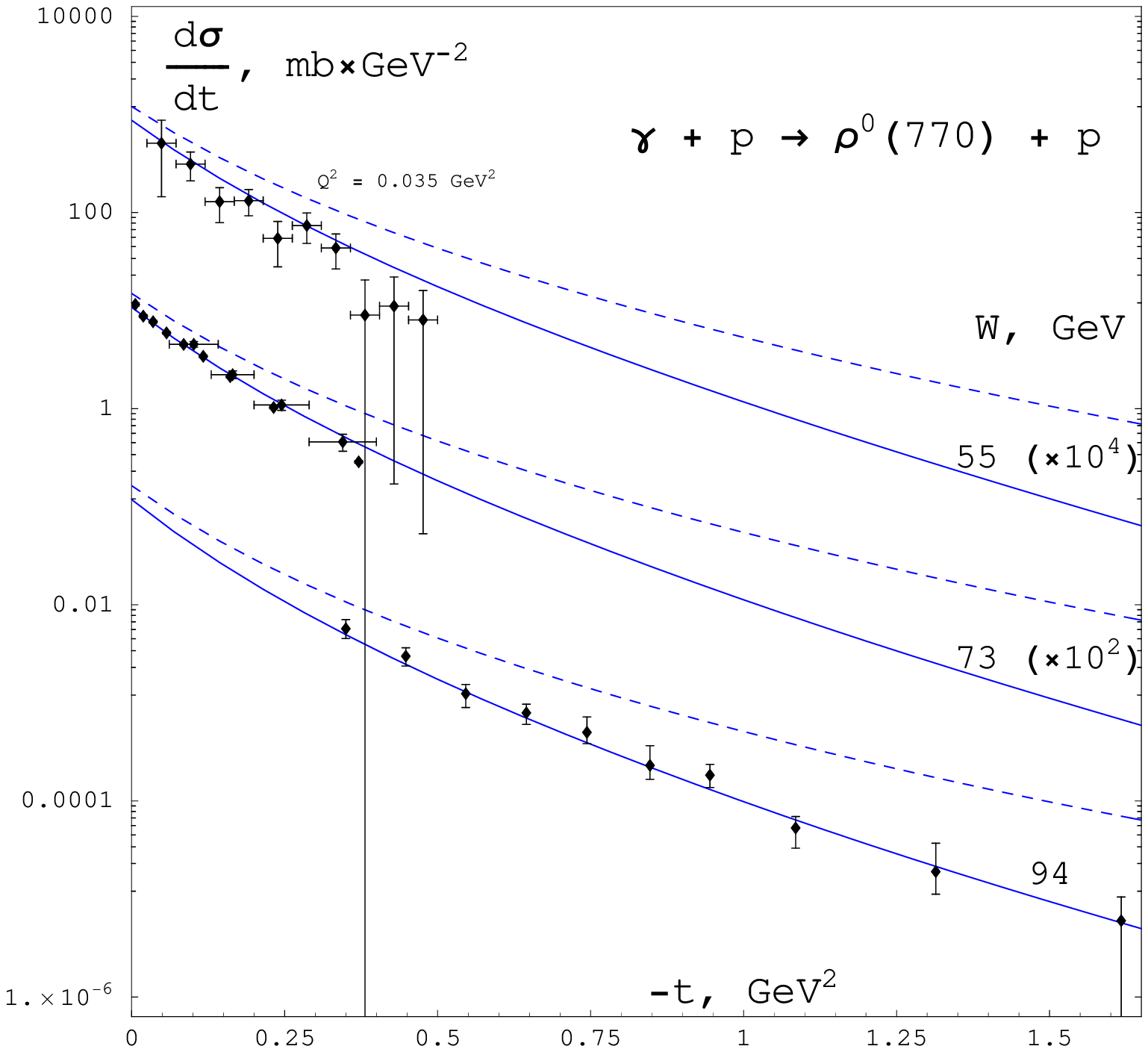}
\vskip -8.2cm
\hskip 8.4cm
\epsfxsize=8.2cm\epsfysize=8.2cm\epsffile{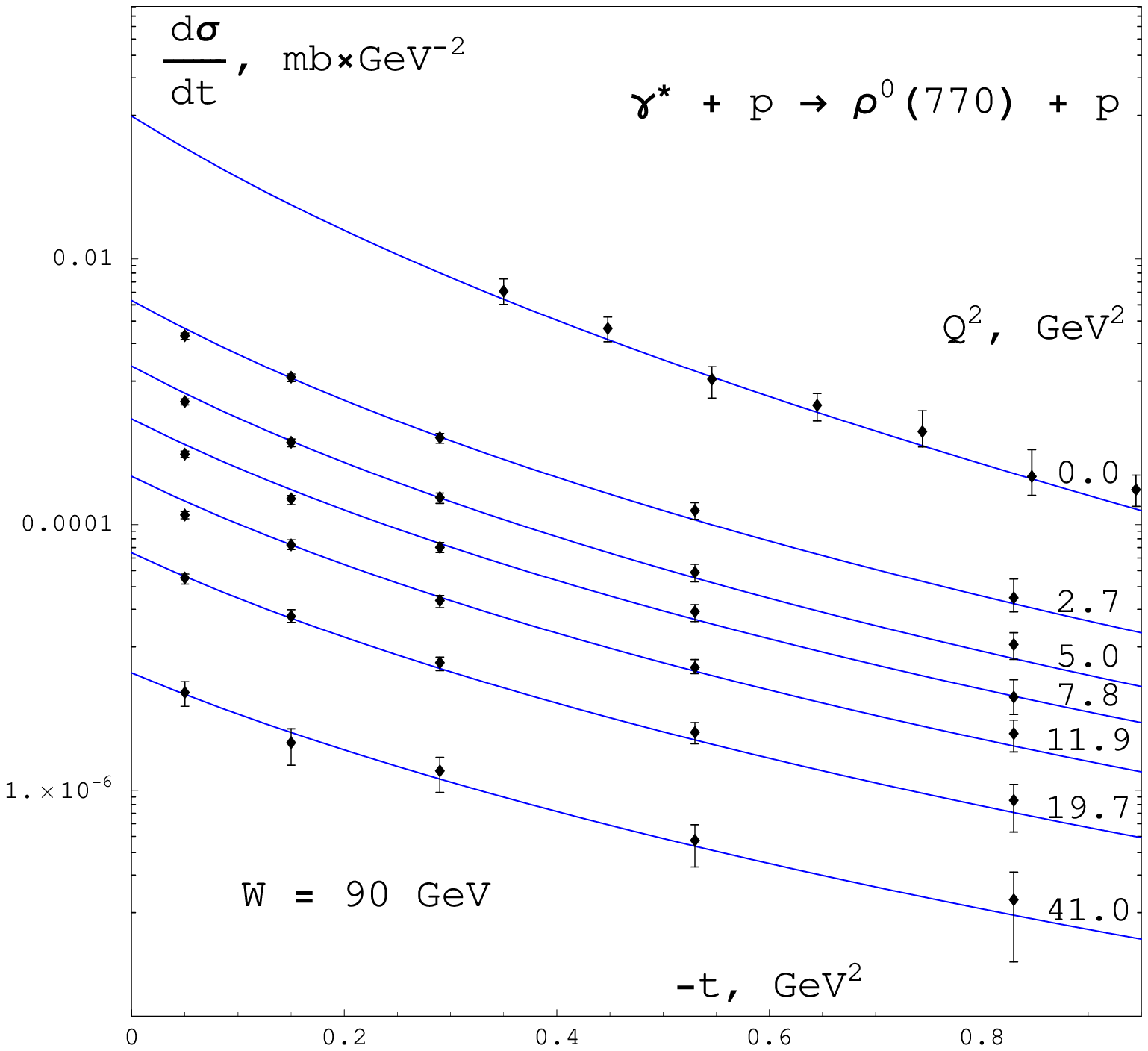}
\end{figure}
\begin{figure}[ht]
\vskip -0.7cm
\epsfxsize=8.2cm\epsfysize=8.2cm\epsffile{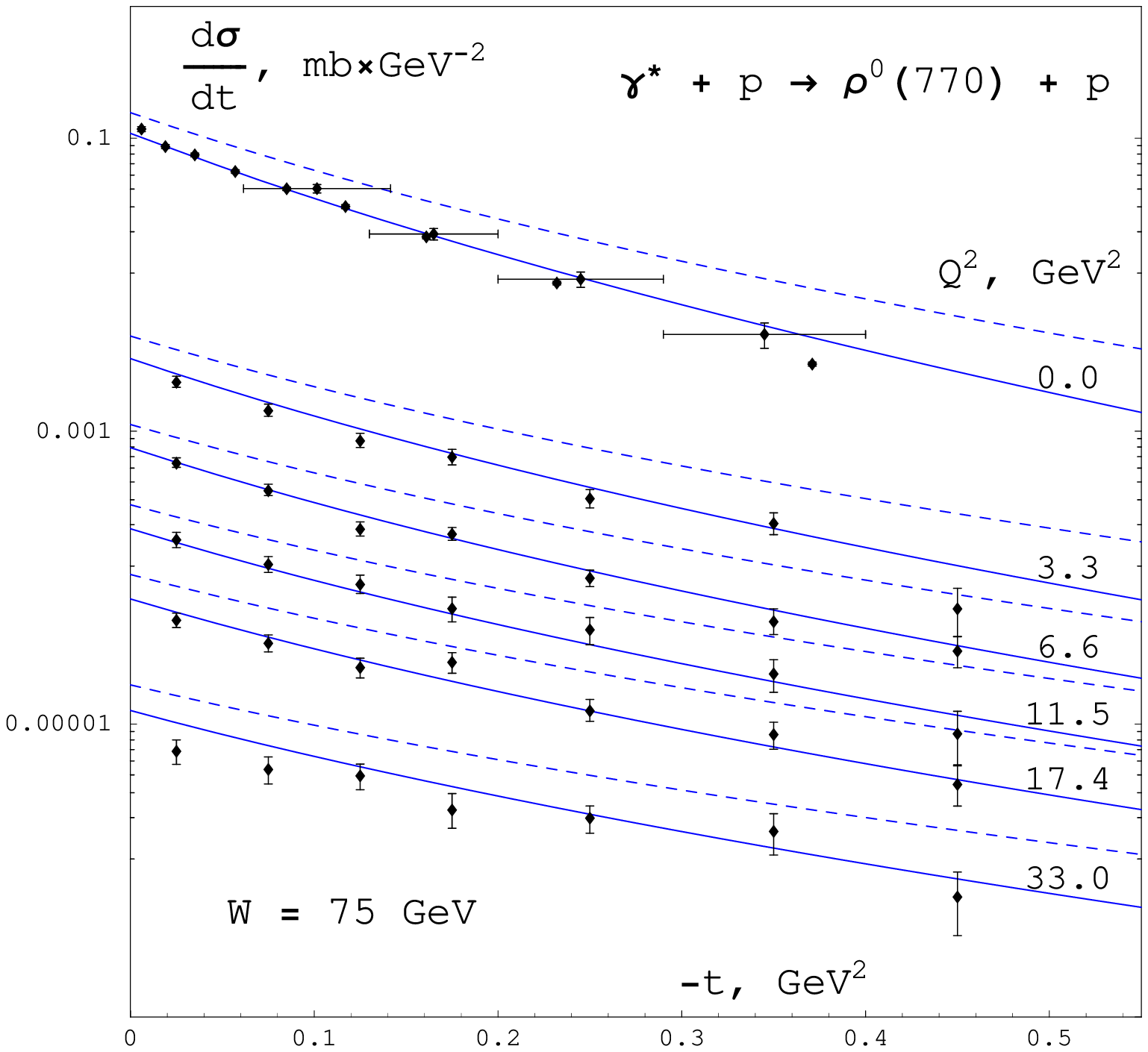}
\vskip -8.2cm
\hskip 8.4cm
\epsfxsize=8.2cm\epsfysize=8.2cm\epsffile{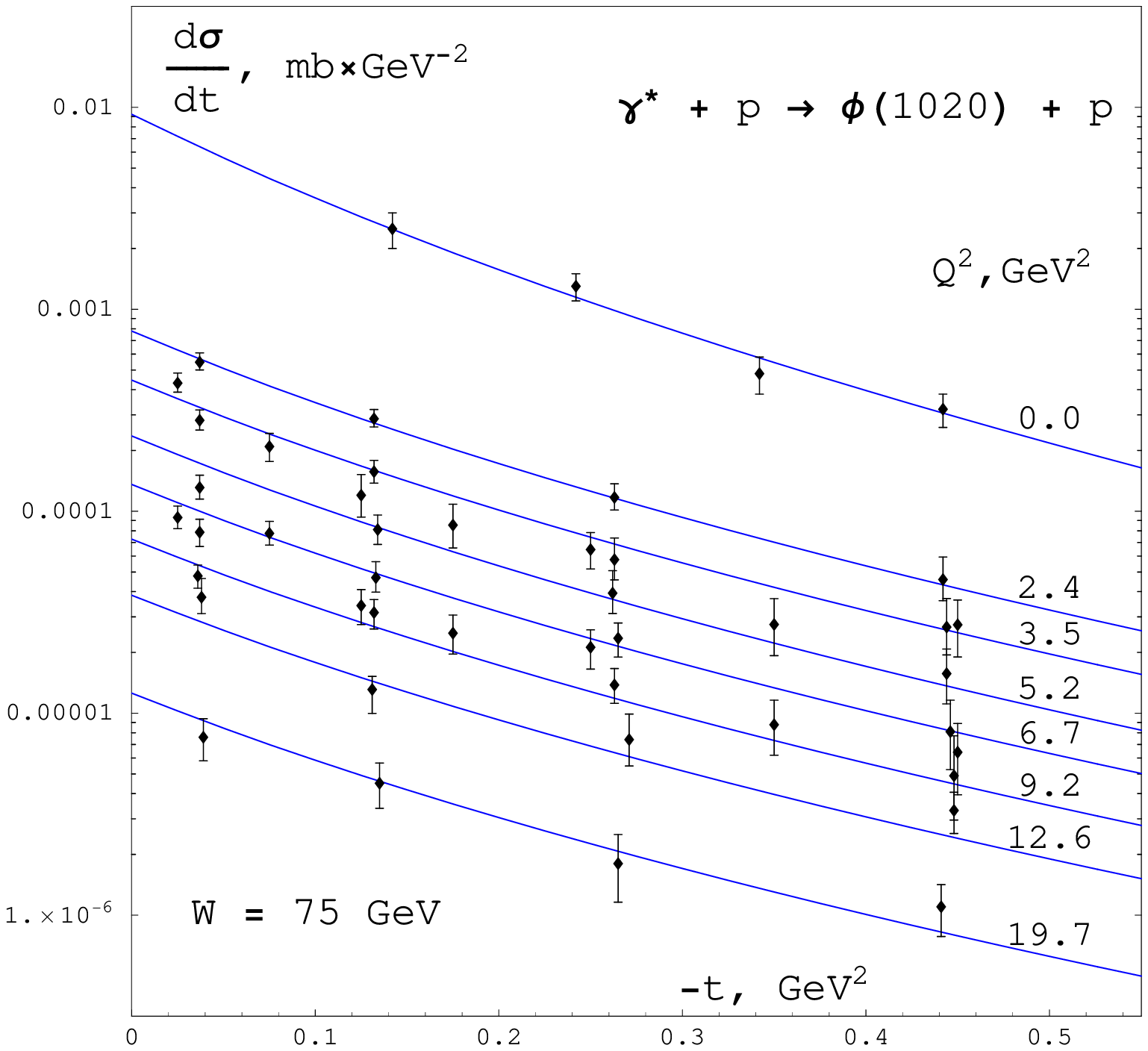}
\caption{The $t$-behavior of the light vector meson electroproduction differential cross-sections at different 
virtualities of the incoming photon. The dashed lines correspond to the Born amplitudes.}
\label{diffli}
\end{figure}

\begin{figure}[ht]
\epsfxsize=8.2cm\epsfysize=8.2cm\epsffile{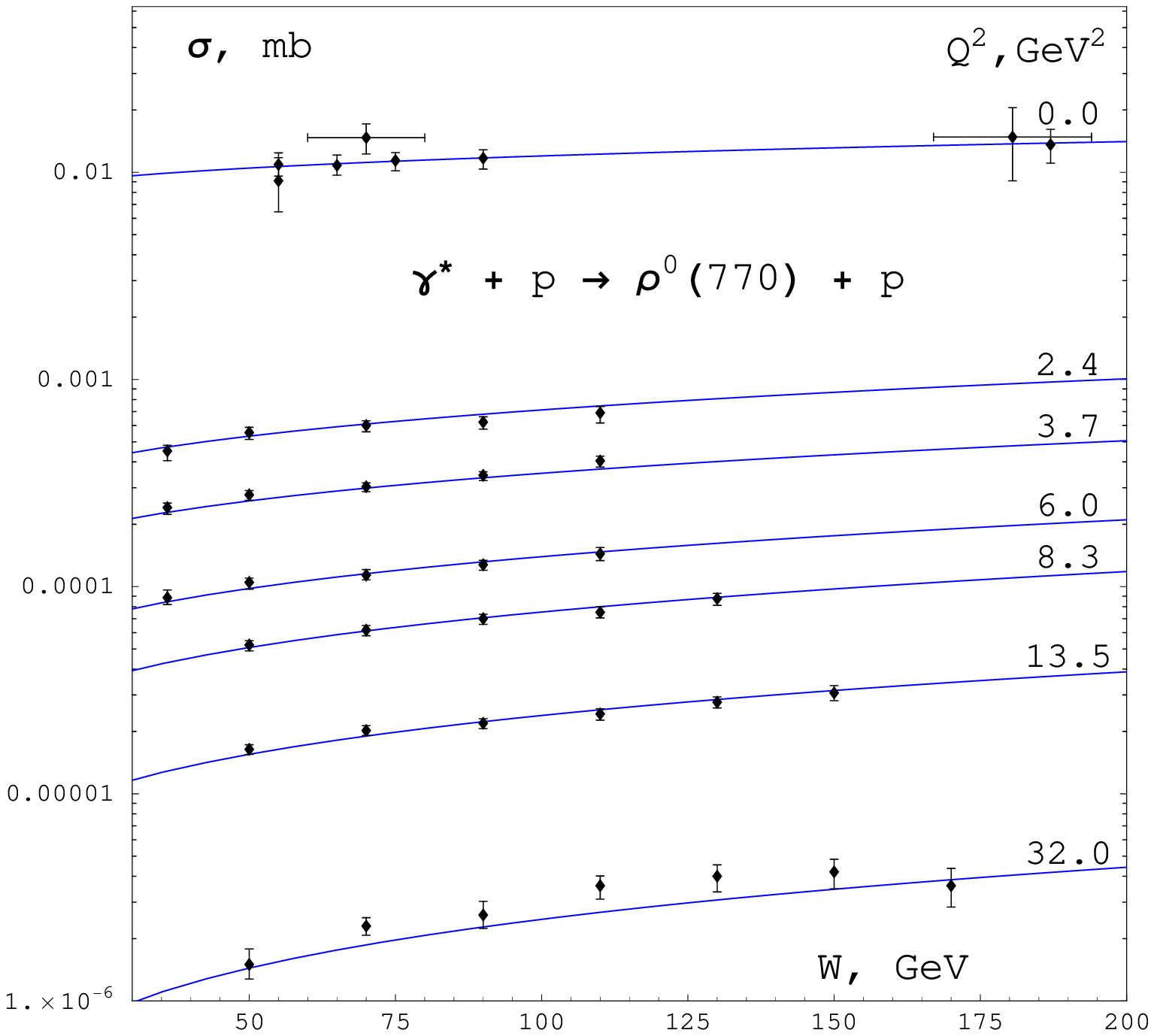}
\vskip -8.2cm
\hskip 8.4cm
\epsfxsize=8.2cm\epsfysize=8.2cm\epsffile{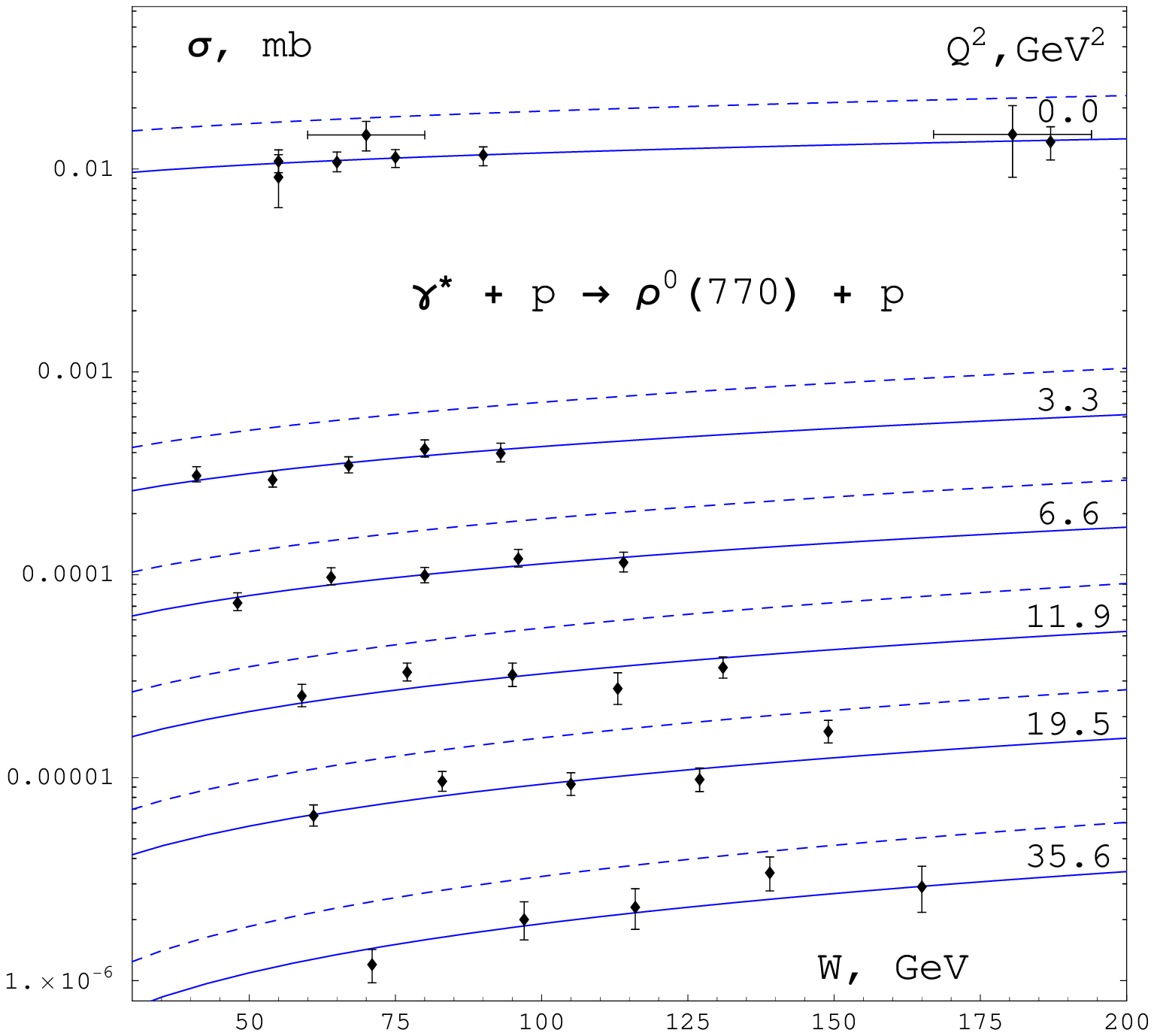}
\end{figure}
\begin{figure}[ht]
\vskip -0.7cm
\epsfxsize=8.1cm\epsfysize=8.1cm\epsffile{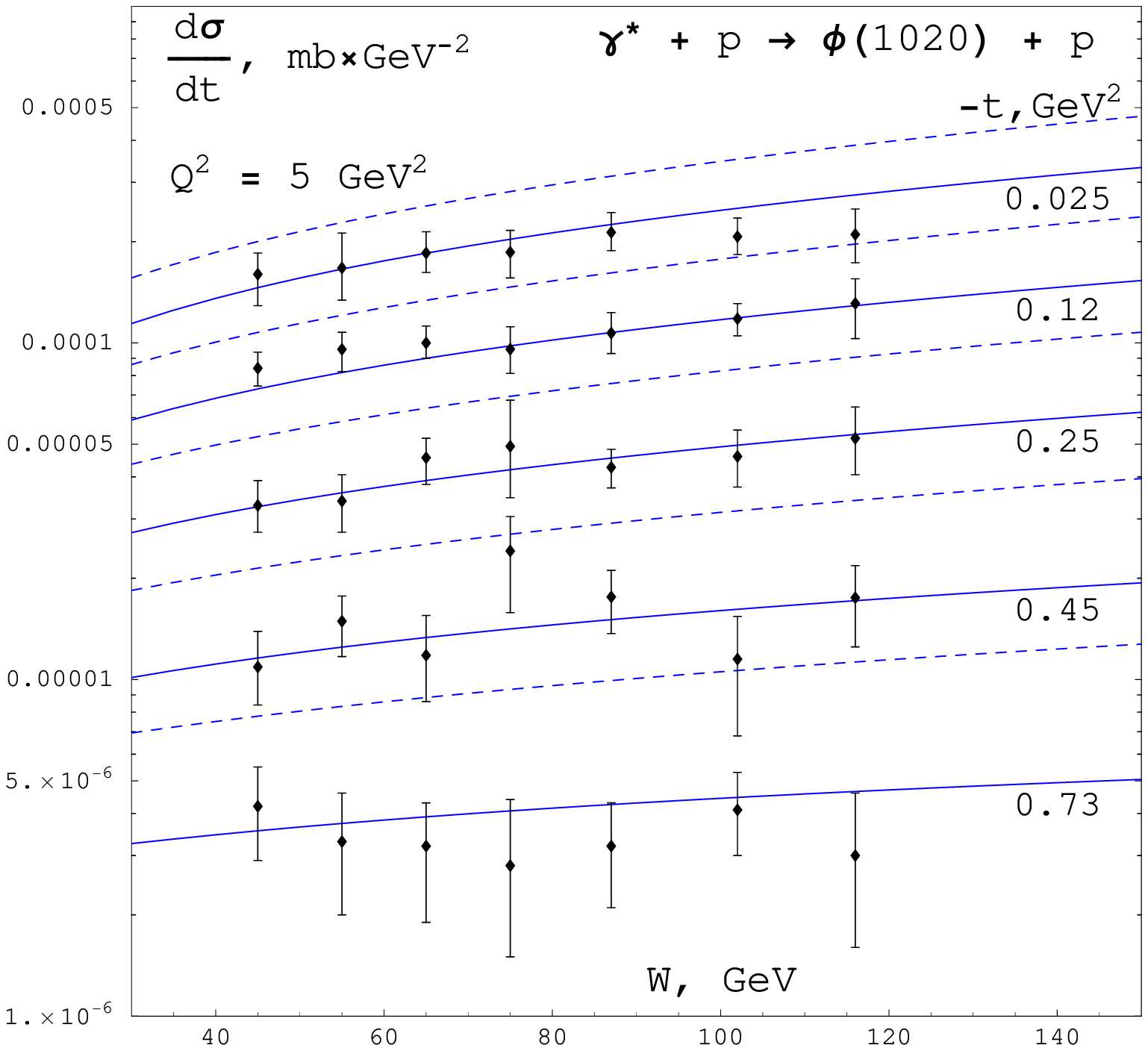}
\vskip -8.2cm
\hskip 8.4cm
\epsfxsize=8.2cm\epsfysize=8.2cm\epsffile{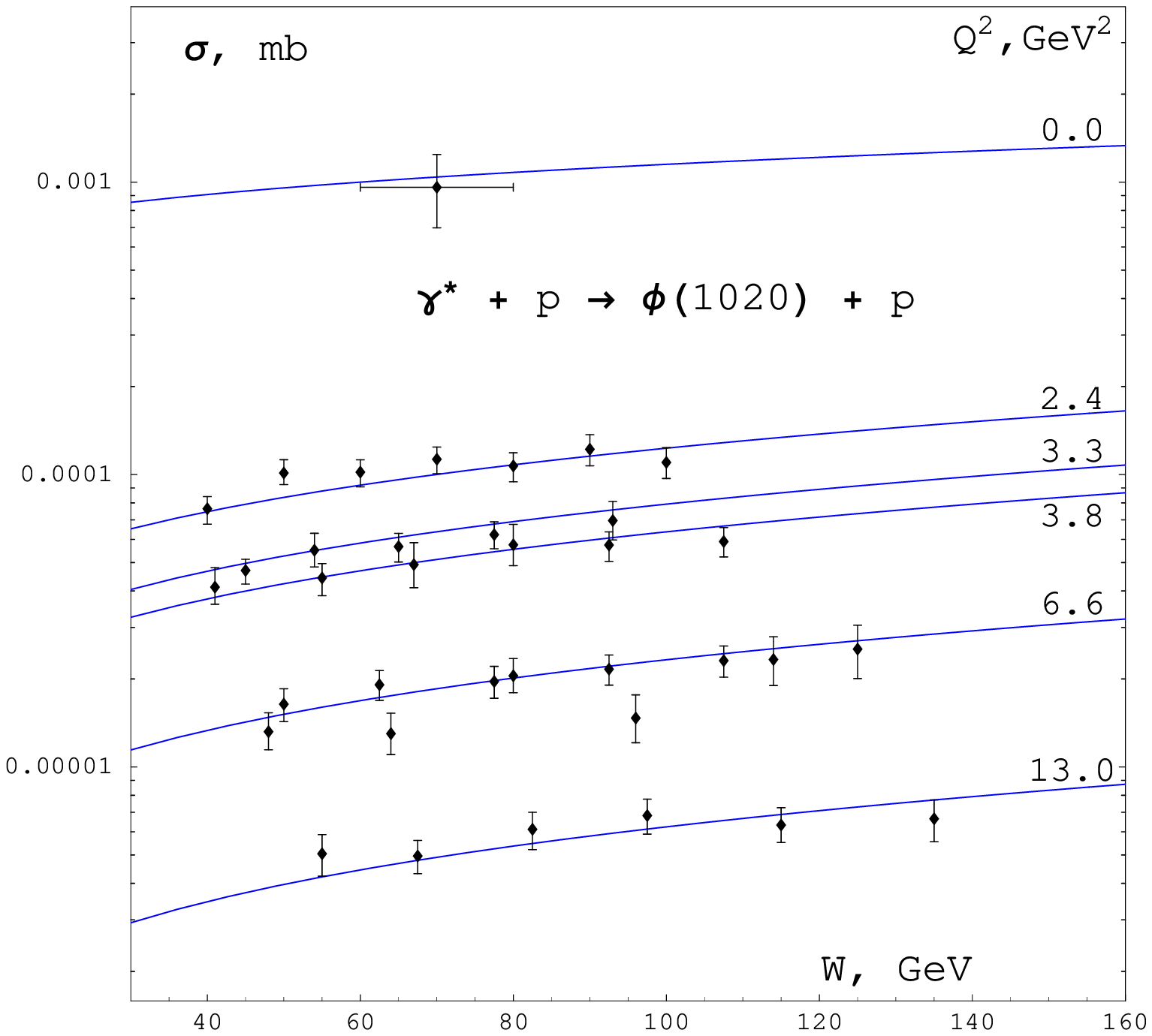}
\caption{The $W$-behavior of the light vector meson electroproduction cross-sections at different 
virtualities of the incoming photon. The dashed lines correspond to the Born amplitudes.}
\label{totli}
\end{figure}

For the electroproduction of $J/\psi(3096)$ \cite{jpsiele,jpsipho} and exclusive production of light vector 
mesons \cite{foto,phiele} we obtain rather slow changing of reggeon form-factors with $Q^2$ growth 
(see Figs. \ref{jpsz},\ref{diffli},\ref{totli}\footnote{Integrated cross-sections are obtained by integration of 
the differential ones 
over $0<-t<1$ GeV$^2$ for heavy mesons and over $0<-t<0.6$ GeV$^2$ for light mesons.} and Tab. \ref{tab3}) 
in accordance with the dimensional counting rules \cite{matveev}.

\begin{figure}[ht]
\vskip -0.3cm
\hskip 4.5cm
\epsfxsize=8.2cm\epsfysize=8.2cm\epsffile{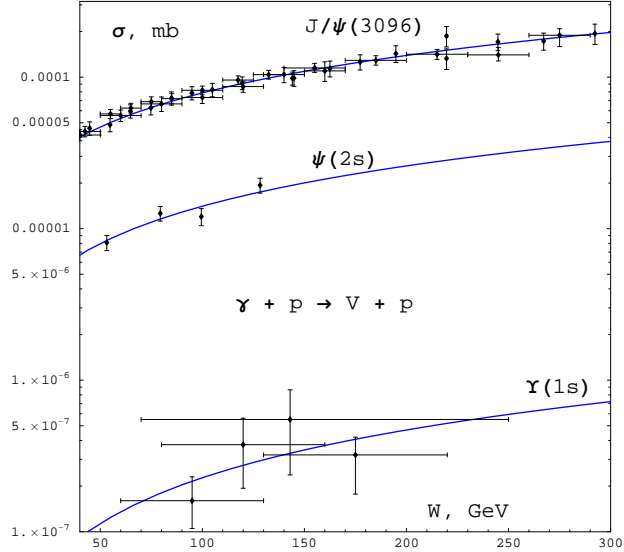}
\caption{High energy behavior of the $\psi(2s)$ and $\Upsilon(1s)$ photoproduction cross-sections.}
\label{ups}
\end{figure}

For the photoproduction of $\psi(2s)$ \cite{psi2} and $\Upsilon(1s)$ \cite{ups} we obtain 
$ \Gamma^{({\rm \psi^*\psi})}_{\rm P}(0)\approx 0.25$, $\Gamma^{({\rm \psi^*\psi})}_f(0)\approx 0.1$ 
and $\Gamma^{({\rm \Upsilon^*\Upsilon})}_{\rm P}(0)\approx 0.08$, $\Gamma^{({\rm \Upsilon^*\Upsilon})}_f(0)\approx 0$ 
(see Fig. \ref{ups}). 

\section*{High energy behavior of $\gamma^*p$ total cross-sections}

In this section we consider the $W$-behavior of the $\gamma^*p$ total cross-sections which are 
closely related to the proton structure function $F_2(x,Q^2)$ \cite{struc}:
\begin{equation}
\label{totga}
\sigma^{\gamma^*p}_{tot}(W^2,Q^2)=\frac{4\pi^2\alpha_e}{Q^4}\frac{Q^2+4 m_p^2x^2}{1-x}F_2(x,Q^2)\;\;\;
\left(x=\frac{Q^2}{W^2+Q^2-m_p^2}\right)\,.
\end{equation} 

Following the VDM the non-flip forward $\gamma^*p$ scattering amplitude can be represented in the form:
\begin{equation}
\label{genvdm2}
T^\lambda_{\gamma^* p\to \gamma^* p}(W^2,t=0,Q^2)=\sum_{V',V}C_{V'}^\lambda(Q^2)
T_{V'^* p\to V^* p}(W^2,t=0,Q^2)C_{V}^\lambda(Q^2)\,.
\end{equation}

Owing to the optical theorem \cite{collins} the total cross-section is proportional to the imaginary part of the forward 
amplitude. From the reggeonic point of view one could single out (at high $W$) the pomeron pole contribution: 
\begin{equation}
\label{totgapom}
\sigma^{\gamma^*p}_{tot}(W^2,Q^2)=\Gamma^{(\gamma^*\gamma^*)}_{\rm P}(Q^2)\Gamma^{({\rm pp})}_{\rm P}(0)
W^{2(\alpha_{\rm P}(0)-1)}+S+A\,,
\end{equation}
where $S$ denotes the sum of other pole contributions, $A$ denotes absorptive corrections, and 
$\Gamma^{(\gamma^*\gamma^*)}_{\rm P}(Q^2)$ is the pomeron form-factor of virtual photon.

Immediately, a question emerges if there exists such a kinematical range at HERA energies where 
$S$ and $A$ are negligible and, hence, the simple pole approximation is valid ($W_0\equiv 1$ GeV):
\begin{equation}
\label{totgapole}
\sigma^{\gamma^*p}_{tot}(W^2,Q^2)\approx\beta(Q^2)\left(\frac{W}{W_0}\right)^{2\delta}\,,
\end{equation}
where $\beta(Q^2)\equiv W_0^{-2}\Gamma^{(\gamma^*\gamma^*)}_{\rm P}(Q^2)\Gamma^{({\rm pp})}_{\rm P}(0)$, 
$\delta\equiv\alpha_{\rm P}(0)-1$.

\begin{figure}[ht]
\vskip -0.7cm
\hskip -0.15cm
\epsfxsize=8.35cm\epsfysize=8.35cm\epsffile{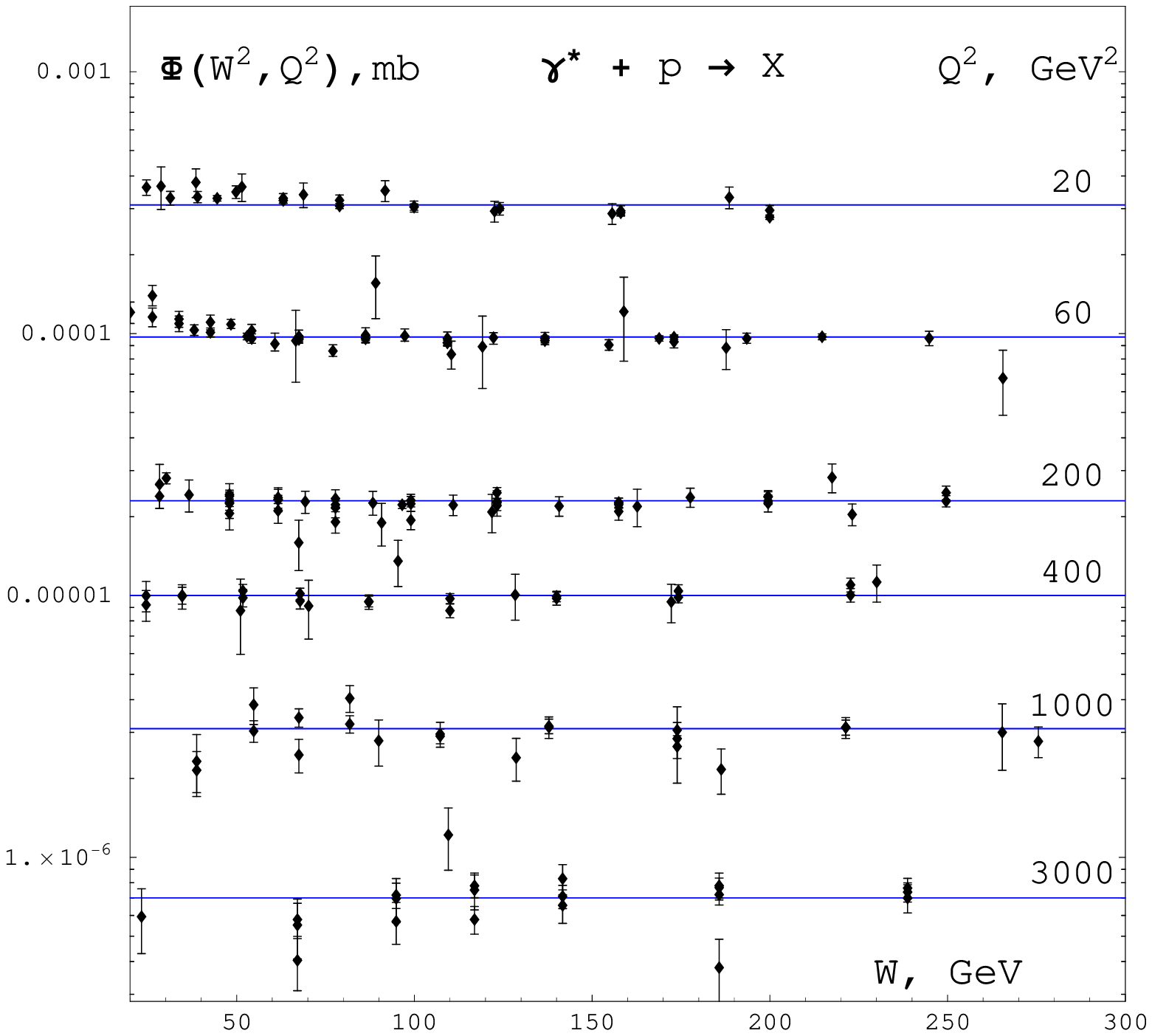}
\vskip -8.4cm
\hskip 8.43cm
\epsfxsize=8.4cm\epsfysize=8.4cm\epsffile{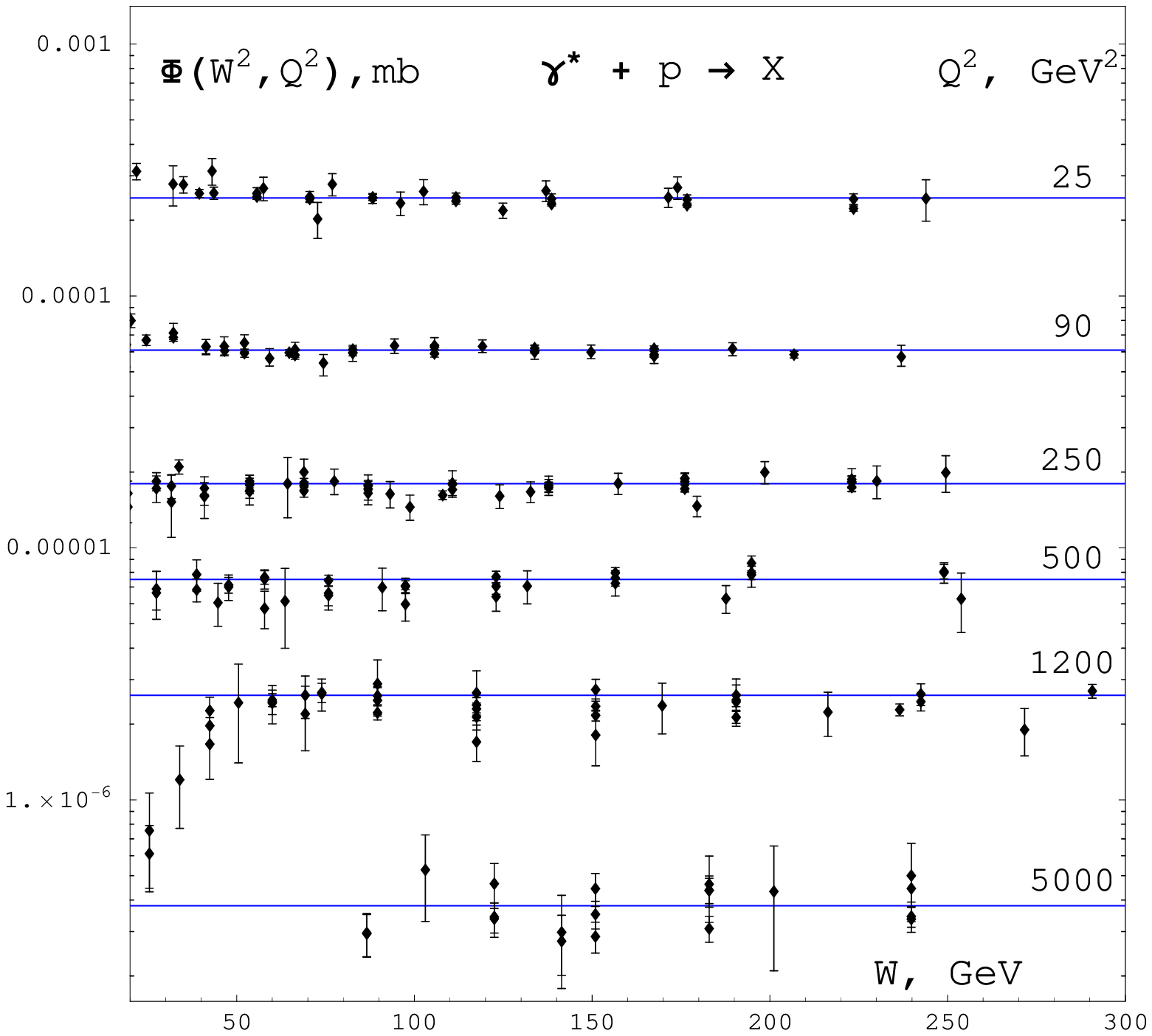}
\caption{The $W$-behavior of the function  $\Phi(W^2,Q^2)\equiv (W_0/W)^{0.62}\sigma_{tot}^{\gamma^*p}(W^2,Q^2)$ 
at different virtualities of the incoming photon.}
\label{stru}
\end{figure}

Such a range exists. At 50 GeV $<W<$ 300 GeV and 25 GeV$^2<Q^2\ll W^2$ the $W$-behavior of the 
$\gamma^*p$ total cross-sections can be well-described by (\ref{totgapole}) with $\delta\approx 0.31$ and 
values of $\beta(Q^2)$ from Tab. \ref{tab4} (see Fig. \ref{stru}). 

\begin{table}[ht]
\begin{center}
\begin{tabular}{|l|l|l|l|l|l|l|l|l|l|l|l|l|l|}
\hline
$Q^2$, GeV$^2$ & 12 & 15 & 20 & 25 & 35 & 45 & 60 & 90 & 120 & 150 & 200 & 250 & 300  \\
\hline
$\beta(Q^2)\cdot 10^4$, mb & 5.0 & 4.1 & 3.1 & 2.45 & 1.7 & 1.32 & 0.97 & 0.61 & 0.44 & 0.33 & 0.23 & 0.18 & 0.145   \\
\hline
\end{tabular}

\vskip 0.3cm

\begin{tabular}{|l|l|l|l|l|l|l|l|l|l|l|l|}
\hline
$Q^2$, GeV$^2$ & 350 & 400 & 500 & 650 & 800 & 1000 & 1200 & 1500 & 2000 & 3000 & 5000  \\
\hline
$\beta(Q^2)\cdot 10^4$, mb & 0.117 & 0.1 & 0.075 & 0.054 & 0.042 & 0.031 & 0.026 & 0.018 & 0.012 & 0.007 & 0.0038   \\
\hline
\end{tabular}
\end{center}
\caption{The $Q^2$-behavior of the pomeron residue for the $\gamma^*p$ forward amplitude.}
\label{tab4}
\end{table}

The slower rising of total cross-sections with the collision energy growth at lower $Q^2$ and $W$ is due to 
the influence of secondary poles and absorptive corrections (the secondary pole terms rise cross-sections 
at lower energies and absorptive corrections drop them at higher energies). The fact that these 
contributions are negligibly small in the above-mentioned kinematical range could be explained by the 
following. Owing to the presence of factors $\frac{M_V^2}{M_V^2+Q^2}$ in coefficients $C_{V}^\lambda(Q^2)$, 
at high photon virtualities the scattering amplitude (\ref{genvdm2}) is dominated by fluctuations to heavy 
vector mesons. But for heavy mesons diffractive scattering the secondary pole terms and the absorptive 
corrections are suppressed by the pomeron pole term due to the much smaller $f_2$-reggeon form-factors 
(in comparison with both the $f_2$-reggeon form-factors of light mesons and the pomeron form-factors) 
and significantly lower values of the pomeron form-factors. This can be traced 
explicitly by the collation of the light meson ($\rho^0(770)$, $\phi(1020)$) versus the heavy meson 
($J/\psi(3096)$, $\psi(2s)$, $\Upsilon(1s)$) exclusive production (see the previous section).

\begin{figure}[ht]
\epsfxsize=8.2cm\epsfysize=8.2cm\epsffile{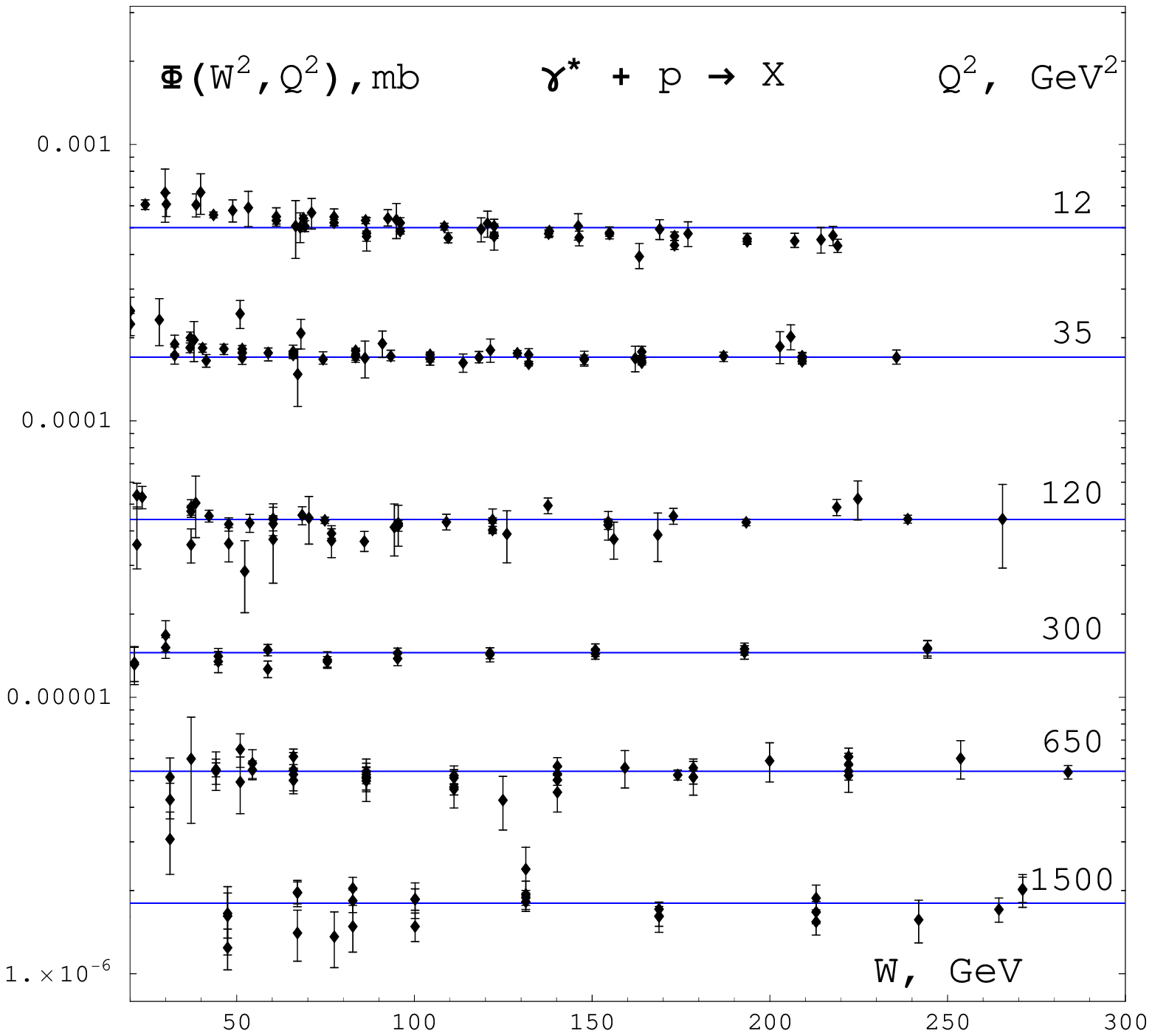}
\vskip -8.3cm
\hskip 8.45cm
\epsfxsize=8.3cm\epsfysize=8.3cm\epsffile{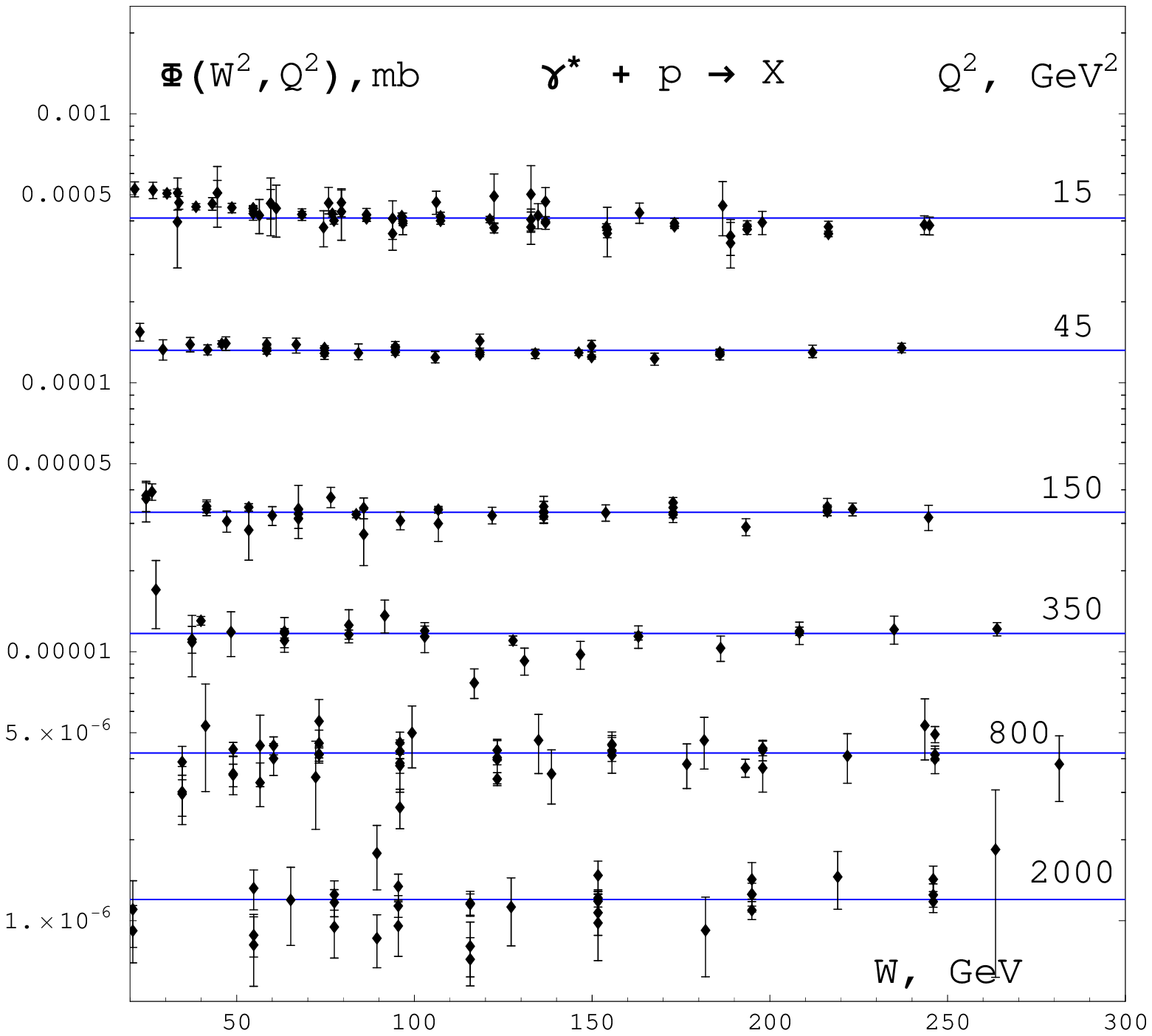}
\end{figure}

After the extraction of $\delta$ from the data on $F_2(x,Q^2)$ and fixing the pomeron intercept 
we obtain (see the previous sections) a consistent phenomenological scheme for various exclusive diffractive 
reactions $2\to 2$ at high energies.

\section*{Discussion}

Now let us turn to the general discussion of the proposed model. 

One of the advantages of Regge-eikonal approach is that it allows explicit taking into account 
absorptive corrections. Above there was demonstrated that absorptive corrections are 
not negligible for elastic nucleon-nucleon scattering and exclusive electroproduction 
of light vector mesons (see dashed lines in Figs. \ref{pp} -- \ref{totli}).

\begin{figure}[ht]
\begin{center}
\epsfxsize=7.7cm\epsfysize=7.7cm\epsffile{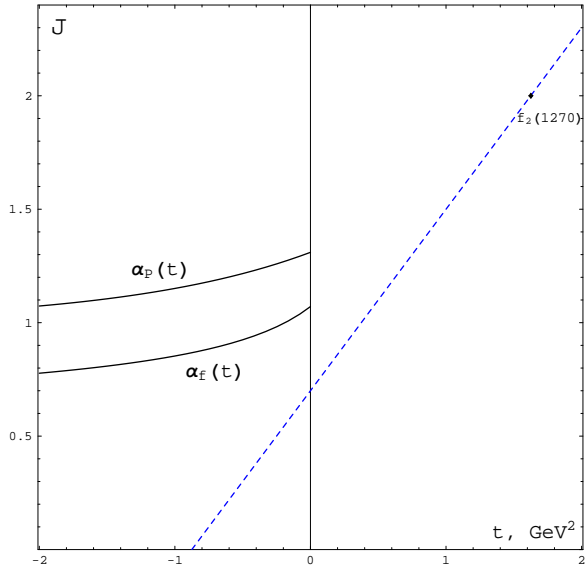}
\caption{Test pomeron and $f_2$-reggeon trajectories. The dashed line, 
$\alpha^{\rm lin}_f(t)=0.7+0.8\,t$, is the Chew-Frautschi plot for the $f_2$-reggeon.}
\label{tra}
\end{center}
\end{figure}

Regge trajectories in our model turn out essentially nonlinear in the diffraction region 
(see Fig. \ref{pp}). At first sight such a picture may seem strange but 
such a nonlinearity is the only way for true Regge trajectories to combine the asymptotic 
behavior (\ref{gluon}), (\ref{meson}) with the approximate large-slope linearity in the 
resonance region\footnote{The resonance masses are determined from equations like 
$\alpha(M^2-{\rm i} M\Gamma)=J$ (here $\alpha(t)$ is some Regge trajectory and $J$ is the 
resonance spin) which do not imply that ${\rm Re}\,\alpha(M^2)=J$ and ${\rm Im}\,\alpha(M^2)=0$. 
The decay widths $\Gamma$ are only few times less than the corresponding masses $M$: for 
example, for $f_2(1270)$-meson $\frac{\Gamma_f}{M_f}\approx 0.15$. Consequently, 
Chew-Frautschi plots should be considered only as very rough approximations to the true Regge 
trajectories in the resonance region.} (observed in the light meson spectroscopy data) 
and the monotony at negative values of the argument (see Fig. \ref{tra}). This monotony itself 
seems quite natural. If ${\rm Im}\,\alpha(t+i0)\ge 0$ 
increases slowly enough at $t\to +\infty$ (for example, not faster than 
$Ct\ln^{-1-\epsilon}t$, $\epsilon>0$), the dispersion 
relations with not more than one subtraction take place, {\it i.e.} 
$$
\alpha(t)=\alpha_0+\frac{t}{\pi}\int_{t_T}^{+\infty}
\frac{{\rm Im}\,\alpha(t'+i0)}{t'(t'-t)}dt'\,.
$$
And if, in addition, we assume that ${\rm Im}\,\alpha(t+i0)\ge 0$ at $t\ge t_T>0$ 
(we would like to point out that these assumptions are strictly fulfilled in the theory 
of perturbations and the theory of potential scattering \cite{collins}) then 
$$
\frac{d^n\alpha(t)}{dt^n}>0\;\;(t<t_T,\;n=1,2,3,...).
$$
Also, using the nonlinear trajectories (instead of linear ones) allows to avoid emerging 
unphysical singularities in the real part of the signature factors at those points where 
Regge trajectories take on a value of negative integers.

The eikonal (\ref{eikphen}) could be called ``minimal'' since it contains contributions from
only those reggeons which are essential for all considered diffractive reactions.
For Regge trajectories and reggeon form-factors we chose parametrizations as simple as possible. 
Expressions (\ref{pomeron}), (\ref{resid}) were not derived from QCD or general principles. 
So, they should be considered not as analytic but only as purely quantitative (test) 
approximations to the true Regge trajectories and reggeon form-factors at negative values 
of the argument. They are not valid at $t>0$. Moreover, we do not expect that exponential 
approximations (\ref{resid}) to reggeon form-factors are valid in the region of transfers essentially 
larger than 1 GeV (exponential form-factors 
correspond to Gaussian distribution of strongly interacting matter in nucleons). 
The imperfectness of these approximations and the neglection of 
secondary reggeons ($\omega$, $\rho$, $a$, {\it etc.}) are the causes 
of systematic deviations of the model curves from the experimental data. 
Therefore, we had to consider only that kinematical range for the elastic nucleon-nucleon 
scattering where the contributions from secondary reggeons are small enough and where the 
non-exponential behavior of reggeon form-factors at high transfers is not essential.

For better description of the considered data sets on nucleon-nucleon scattering and for 
extension of the model to higher transfers one should use more complicated expressions 
for the reggeon form-factors. Also, for extension to lower energies and for description of 
the difference between $pp$ and $\bar p p$ cross-sections it is necessary to include 
secondary reggeons ($\omega$, $\rho$, $a$, {\it etc.}) and, possibly, odderon 
(which could have some influence on the behavior in the dip region).
But in this job we deliberately restricted ourselves by the simplest eikonal and the simplest 
test parametrizations to make the main conclusions more transparent:
\begin{itemize}
\item For various diffractive processes there exist wide kinematical ranges where the 
cross-sections are dominated by only two reggeons, the pomeron and the $f_2$-reggeon.
\item Regge trajectories are universal in all diffractive reactions\footnote{This universality 
is closely related to the fact that Regge trajectories are analytic 
({\it i.e.} unique) continuations of observable spectra of resonances.} and 
essentially nonlinear (such a nonlinearity does not contradict to both the 
phenomenology and the QCD asymptotic relations).
\item The pomeron trajectory has intercept $\approx 1.31$ (which could be extracted from 
the data on the proton structure function $F_2(x,Q^2)$) 
and corresponds to the so-called ``hard'' 
pomeron frequently mentioned in literature. The $f_2$-reggeon has intercept $\approx 1.07$ and corresponds 
to the so-called ``soft'' pomeron.
\item Different cross-section growth rates for various reactions (at high energies) are due to different 
relative contributions of the $f_2$-reggeon to the eikonal.
\end{itemize}

Among other models for elastic diffractive scattering 
\cite{martynov,prokudin,soffer,menon,martynov2,nicolescu} and 
exclusive electroproduction of vector mesons 
\cite{predazzi,fiore,motyka,soyez,donnachie} the proposed phenomenological scheme stands 
out due to its salient simplicity and physical clearness and, also, due to the fact that it is 
the only model which allows to give a simultaneous qualitative description to both ``soft'' and 
``hard'' exclusive 
diffractive processes. The range of validity of the model is wide enough to give a good ground 
for making well-reasoned predictions for elastic diffraction at LHC energies (see Fig. \ref{pp}). 
The TOTEM measurements \cite{totem} on $pp$ total cross-sections and angular distributions should 
discriminate among different models. 

The obtained approximations to leading vacuum Regge trajectories and reggeon form-factors of the proton 
could be used under considering more complicated (than $2\to 2$) diffractive reactions: 
single diffraction ($p+p\to p+X$ or $\bar p+p\to \bar p+X$), central exclusive diffractive production 
of the Higgs boson ($p+p\to p+H+p$), {\it etc.}

\section*{Acknowledgments}

The author is grateful to V.A. Petrov for encouragement and numerous discussions and 
A.K. Likhoded, G.P. Pronko, V.V. Kiselev, and Yu.F. Pirogov for discussion and useful criticism.

\section*{Appendix. Derivation of the eikonal Regge approximation.}

The eikonal representation of the scattering amplitude itself does not yield any progress 
in solving the problem since it is reduced to the replacement of the unknown function of 
two variables, $T(s,t)$, to another one, $\delta(s,t)$, without any specification of the 
functional form of $\delta(s,t)$. The key assumption is that the eikonal is proportional 
(with high accuracy) to the effective relativistic (quasi-)potential of hadronic 
interaction. According to the Van Hove interpretation \cite{vanhove} of the relativistic 
(quasi-)potential as the ``sum'' over all single-particle exchanges in the 
$t$-channel, the eikonal can be represented in the form
$$
\delta^{(f_1,f_2)}(s,t) = \sum_{j=0}^{\infty}\sum_{m_j} 
J^{(f_1,j,m_j)}_{\alpha_1...\alpha_j}(p_1,\Delta)
\frac{D_{(j,m_j)}^{\alpha_1...\alpha_j,\beta_1...\beta_j}
(\Delta)}{m_j^2-\Delta^2}J^{(f_2,j,m_j)}_{\beta_1...\beta_j}(p_2,\Delta)\,,
\eqno{(\rm A.1)}
$$
(here $\frac{D_{(j,m_j)}^{\alpha_1...\alpha_j,\beta_1...\beta_j}}{m_j^2-\Delta^2}$ is the 
propagator of the spin-$j$ particle with mass $m_j$, $J^{(f,j,m_j)}_{\alpha_1...\alpha_j}$ 
is the hadronic current (index $f$ denotes the kind of scattering particle), $\Delta$ 
is the transferred 4-momentums, $t = \Delta^2$, $p_1$ and $p_2$ are 4-momentumes of the 
incoming hadrons, $s=(p_1+p_2)^2$, the symbol $\sum_{m_j}$ denotes summing over all 
spin-$j$ particles with different masses, which, in what further, will be transformed into 
summing over reggeons). We impose the following constraints on the general dependence of 
hadronic currents on $\Delta$: symmetry with respect to all $\alpha_k$, transversality 
with respect to $\Delta_{\alpha_k}$ ($k = 1,...,j$), and tracelessness with respect to 
any pair of Lorentz indices. The first two conditions yield 
$$
J^{(f,j,m_j)}_{\alpha_1...\alpha_j}(p,\Delta) = 
\sum_{k=0}^{\left[\frac{j}{2}\right]}\Gamma^{(f,j,m_j)}_k(p^2,\Delta^2,(p\Delta))
\sum G_{\alpha_{\mu_1}\alpha_{\mu_2}}...G_{\alpha_{\mu_{2k-1}}\alpha_{\mu_{2k}}} 
P_{\alpha_{\mu_{k+1}}}...P_{\alpha_{\mu_j}},
\eqno{(\rm A.2)}
$$
where $\Gamma^{(f,j,m_j)}_k(p^2,\Delta^2,(p\Delta))$ are some scalar functions, $P_{\alpha}\equiv 
\frac{p_{\alpha} - \frac{p\Delta}{\Delta^2}\Delta_{\alpha}}{\sqrt{p^2-\frac{(p\Delta)^2}{\Delta^2}}}$, 
$G_{\alpha\beta}\equiv -g_{\alpha\beta} + \frac{\Delta_{\alpha}\Delta_{\beta}}{\Delta^2}$, 
and the inner sum is over all nonequivalent permutations of Lorentz indices (the total number of 
terms is $\frac{j!}{(2k)!!(j-2k)!}$). Taking into account that 
$$
g^{\alpha\beta}G_{\alpha\beta} = -3\,,\;\;\;P_{\alpha}P^{\alpha} = 1\,,\;\;\;
g^{\alpha\beta}G_{\alpha\gamma}G_{\beta\delta} = -G_{\gamma\delta}\,,\;\;\;
G_{\alpha\beta}P^{\beta} = -P_{\alpha}\,,
\eqno{(\rm A.3)}
$$
the tracelessness condition results in the recurrent relations 
$$
\Gamma^{(f,j,m_j)}_k(p^2,\Delta^2,(p\Delta)) = 
\frac{\Gamma^{(f,j,m_j)}_{k-1}(p^2,\Delta^2,(p\Delta))}{2(j-k)+1}\,,
\eqno{(\rm A.4)}
$$
and, consequently, yields
$$
J^{(f,j,m_j)}_{\alpha_1...\alpha_j}(p,\Delta) = \Gamma^{(f,j,m_j)}_0(p^2,\Delta^2,(p\Delta))\times
\eqno{(\rm A.5)}
$$
$$
\times\sum_{k=0}^{\left[\frac{j}{2}\right]}\frac{(2(j-k)-1)!!}{(2j-1)!!}
\sum G_{\alpha_{\mu_1}\alpha_{\mu_2}}...G_{\alpha_{\mu_{2k-1}}\alpha_{\mu_{2k}}} 
P_{\alpha_{\mu_{k+1}}}...P_{\alpha_{\mu_j}}\,.
$$
Substituting (\rm A.5) into (\rm A.1) and taking into account the transversality and the 
tracelessness of hadronic currents, we come to the following expression for the eikonal:
$$
\delta^{(f_1,f_2)}(s,t) = \sum_{j=0}^{\infty}\sum_{m_j}
\frac{\gamma^{(f_1,f_2,j,m_j)}(\Delta^2)}{m_j^2-\Delta^2}\frac{2^j(j!)^2}{(2j)!}
P_j\left(\frac{p_1p_2-\frac{(p_1\Delta)(p_2\Delta)}{\Delta^2}}
{\sqrt{(p_1^2-\frac{(p_1\Delta)^2}{\Delta^2})(p_2^2-\frac{(p_2\Delta)^2}{\Delta^2})}}\right),
\eqno{(\rm A.6)}
$$
where $P_j(x)$ are Legendre polynomials of power $j$ and 
$$
\gamma^{(f_1,f_2,j,m_j)}(\Delta^2)\equiv \Gamma^{(f_1,j,m_j)}_0(p_1^2,\Delta^2,(p_1\Delta))
\Gamma^{(f_2,j,m_j)}_0(p_2^2,\Delta^2,(p_2\Delta))\,.
\eqno{(\rm A.7)}
$$
For elastic scattering we have  
$$
(p_1-\Delta)^2 = p_1^2 = m_1^2\,,\;\;\; (p_2+\Delta)^2 = p_2^2 = m_2^2\,,
\eqno{(\rm A.8)}
$$
and, so, using the kinematical relations 
$$
p_{1,2}\Delta = \pm\frac{\Delta^2}{2}\,,\;\;\;\;
p_1p_2 = \frac{s - m_1^2 - m_2^2}{2}\,,\;\;\;\;\Delta^2\equiv t\,,
\eqno{(\rm A.9)}
$$
we can simplify the expression for the eikonal (from now on, indices $f_1$ and $f_2$, 
denoting the kinds of scattering particles, will be omitted):
$$
\delta(s,t) = \sum_{j=0}^{\infty}\sum_{m_j}\frac{\gamma^{(j,m_j)}(t)}{m_j^2-t}
\frac{2^j(j!)^2}{(2j)!}P_j\left(\frac{s-m_1^2-m_2^2+\frac{t}{2}}
{2\sqrt{(m_1^2-\frac{t}{4})(m_2^2-\frac{t}{4})}}\right)\,.
\eqno{(\rm A.10)}
$$
Here we divide the sum over $j$ in the right-hand side of (\rm A.10) into two sums 
over even and odd $j$:
$$
\delta(s,t) = \delta^+(s,t) + \delta^-(s,t) = 
\sum_{j=0}^{\infty}
\sum_{\eta = \pm 1}\sum_{m_{(\eta)j}}\frac{\eta+e^{-i\pi j}}{2}(-1)^j
\frac{\gamma^{(\eta,j,m_{(\eta)j})}(t)}{m_{(\eta)j}^2-t}
\times
\eqno{(\rm A.11)}
$$
$$
\times
\frac{2^j(j!)^2}{(2j)!}P_j\left(\frac{s-m_1^2-m_2^2+\frac{t}{2}}
{2\sqrt{(m_1^2-\frac{t}{4})(m_2^2-\frac{t}{4})}}\right)\,.
$$
Each of sequences $\gamma^{(\eta,j,m_{(\eta)j})}(t)$, $m_{(\eta)j}^2$ ($j = 0,2,4,...,2n,...$ 
at $\eta = +1$ and $j = 1,3,4,...,2n-1,...$ at $\eta = -1$) separately satisfies the conditions 
of the Carlson theorem \cite{collins,carlson} which states that if for some analytic and regular 
at ${\rm Re}\,x\ge 0$ function $f(x)$ the inequality $f(x)<e^{k|x|}$ at some $k<\pi$ 
is valid, then this function is determined unilocally by its values at integer $x$.

Now we make a key assumption about the possibility of unilocal (under the Carlson theorem) analytic 
continuation of (\rm A.11) into the region of complex $j$ (the Regge hypothesis \cite{collins}).
It implies that $m_{(\eta)j}^2$ and $\gamma^{(\eta,j,m_{(\eta)j})}(t)$ in (\rm A.11) are 
the values of analytic (holomorphic with respect to the complex variable $j$) functions 
for integer non-negative values of $j$. We denote these functions by $m_{\eta}^2(j)$ and 
$\gamma^{(\eta)}(t,j,m_{\eta}^2(j))$, respectively. Via the Sommerfeld–Watson transform 
\cite{collins,zommer} we replace the sum over $j$ in (\rm A.11) by the integral over the contour $C$ 
(on the complex plane of the variable $j$) encircling the real positive half-axis including the point 
$j = 0$ in such a way that the half-axis is on the right:
$$
\delta(s,t) = \frac{1}{2i}\oint_C\frac{dj}{\sin(\pi j)}\sum_{\eta = \pm 1}
\sum_{m_{\eta}}\left(-\frac{\eta+e^{-i\pi j}}{2}\right)
\frac{\gamma^{(\eta)}(t,j,m_{\eta}^2(j))}
{m_{\eta}^{2}(j)-t}\times
\eqno{(\rm A.12)}
$$
$$
\times\frac{2^j\Gamma^2(j+1)}{\Gamma(2j+1)}
P\left(j\;,\;\frac{s-m_1^2-m_2^2+\frac{t}{2}}
{2\sqrt{(m_1^2-\frac{t}{4})(m_2^2-\frac{t}{4})}}\right)
$$
(here $P(j,x)$ is the analytic continuation of Legendre polynomials $P_j(x)$ to the region of complex 
$j$ and $\Gamma(j)$ is the Euler gamma-function). Since (according to our assumption) the unique
sources of singularities of the integrand in the region ${\rm Re}\;j >-\frac{1}{2}$ are the zeros of 
the functions $\sin(\pi j)$ and $m_{\eta}^2(j)-t$, then, by deforming the contour $C$ and passing to
the contour parallel to the imaginary axis, ${\rm Re}\;j = -\frac{1}{2}$, we obtain 
$$
\delta(s,t) = \frac{1}{2i}\int_{-\frac{1}{2}-i\infty}^{-\frac{1}{2}+i\infty}
\frac{dj}{\sin(\pi j)}\sum_{\eta = \pm 1}\sum_{m_{\eta}}
\left(-\frac{\eta+e^{-i\pi j}}{2}\right)\frac{\gamma^{(\eta)}(t,j,m_{\eta}^2(j))}
{m_{\eta}^{2}(j)-t}\times
$$
$$
\times\frac{2^j\Gamma^2(j+1)}{\Gamma(2j+1)}
P\left(j\;,\;\frac{s-m_1^2-m_2^2+\frac{t}{2}}
{2\sqrt{(m_1^2-\frac{t}{4})(m_2^2-\frac{t}{4})}}\right)+
\eqno{(\rm A.13)}
$$
$$
+\sum_{\eta = \pm 1}\sum_n\left(-\frac{\eta+e^{-i\pi \alpha_n^{(\eta)}(t)}}
{\sin(\pi\alpha_n^{(\eta)}(t))}\right)\frac{d\alpha_n^{(\eta)}(t)}{dt}\,
\frac{\pi \gamma^{(\eta)}(t,\alpha_n^{(\eta)}(t),t)}{2}\times
$$
$$
\times\frac{2^{\alpha_n^{(\eta)}(t)}\Gamma^2(\alpha_n^{(\eta)}(t)+1)}
{\Gamma(2\alpha_n^{(\eta)}(t)+1)}
P\left(\alpha_n^{(\eta)}(t)\;,\;\frac{s-m_1^2-m_2^2+\frac{t}{2}}
{2\sqrt{(m_1^2-\frac{t}{4})(m_2^2-\frac{t}{4})}}\right)\,,
$$
where the functions $\alpha_n^{(\eta)}(t)$ are the roots of the equations $m_{\eta}^2(j)-t = 0$ and, 
thus, correspond to the poles of the eikonal in the region of complex $j$. These poles are called Regge 
poles, and the functions $\alpha_n^{(\eta)}(t)$ are called Regge trajectories ($C$-even at $\eta=+1$ 
and $C$-odd at $\eta=-1$). 

For $s\gg m_1^2+m_2^2-\frac{t}{2}$ we can neglect the contribution of the background integral 
and the Legendre polynomials can be described by the leading terms of the expansion. Therefore, 
it is convenient to introduce new functions 
$$
\beta_n^{(\eta,s_0)}(t)\equiv\frac{d\alpha_n^{(\eta)}(t)}{dt}\,
\frac{\pi \gamma^{(\eta)}(t,\alpha_n^{(\eta)}(t),t)}{2}
\left(\frac{s_0}{2\sqrt{(m_1^2-\frac{t}{4})
(m_2^2-\frac{t}{4})}}\right)^{\alpha_n^{(\eta)}(t)}\,,
\eqno{(\rm A.14)}
$$
where $s_0$ is any scale determined {\it a priori} (for example, $s_0 = 1$ GeV$^2$). 
Note, that functions $\beta_n^{(\eta,s_0)}(t)$ depend on the chosen scale $s_0$ and can be 
factorized into two factors (see A.7) corresponding to the reggeon form-factors of the 
scattering particles. In the limit of high energies we obtain 
$$
\delta(s,t) = \sum_n\left(i+{\rm tg}\frac{\pi(\alpha_n^+(t)-1)}{2}\right)
\Gamma_n^{(1)+}(t)\Gamma_n^{(2)+}(t)\left(\frac{s}{s_0}\right)^{\alpha_n^+(t)}\mp
\eqno{(\rm A.15)}
$$
$$
\mp\sum_n\left(i-{\rm ctg}\frac{\pi(\alpha_n^-(t)-1)}{2}\right)
\Gamma_n^{(1)-}(t)\Gamma_n^{(2)-}(t)\left(\frac{s}{s_0}\right)^{\alpha_n^-(t)}\,,
$$
where the sign ``$-$'' (``$+$'') before $C$-odd contributions corresponds to the 
particle-particle (particle-antiparticle) scattering and $\Gamma_n^{(i)\pm}(t)$ 
are reggeon form-factors of the scattered particles. 

The last formula\footnote{It is valid, also, for those single-particle exchanges for 
which (\rm A.8) is violated, {\it i.e.}, as for the inelastic scattering $2\to 2$ so 
for reactions with off-shell particles.}, together
with the eikonal representation of the scattering amplitude (\ref{eikrepr}), is the
essence of the Regge-eikonal model \cite{collins}.

\end{document}